\documentclass{aa}
\input{epsf.sty}
\usepackage{graphicx}
\usepackage{natbib}
\def\about{$\simeq$}

\newcommand{\Al}{$^{26\!}$Al\ }
\newcommand{\ms}{$M_\odot$\ }
\def\beq{\begin{equation}}
\def\enq{\end{equation}}

\begin{document}

\title{Spectral and intensity variations of Galactic \Al emission}

\institute{Max-Planck-Institut f\"{u}r extraterrestrische Physik,
Postfach 1312, 85741 Garching, Germany \and National Astronomical
Observatories, Chinese Academy of Sciences, Beijing 100012, China
\and AstroParticule et Cosmologie (APC), 11 Place Marcelin
Berthelot, 75231 Paris, France \and Centre d'Etude Spatiale des
Rayonnements, B.P.N$^\circ$ 4346, 31028 Toulouse Cedex 4, France
\and ESA-ESTEC, Keplerlaan 1, 2201 AZ Noordwijk, The Netherlands
 }

\author{W. Wang\inst{1,2}, M.G.
Lang\inst{1}, R. Diehl\inst{1}, H. Halloin\inst{3}, P.
Jean\inst{4}, J. Kn\"odlseder\inst{4}, K. Kretschmer\inst{1}, P.
Martin\inst{4}, P. Roques\inst{4}, A.W. Strong\inst{1}, C.
Winkler\inst{5}, and X.L. Zhang\inst{1} }

\offprints{W. Wang, wwang@mpe.mpg.de}
\date{Received }

\authorrunning{W. Wang et al.}
\titlerunning{$^{26}$Al emission along the Galactic plane}

\abstract {}
  {Gamma-ray line emission from the radioactive decay of \Al
reflects nucleosynthesis in massive stars and supernovae. We use
INTEGRAL \Al measurements to characterize the distribution and
characteristics of \Al source regions throughout the Galaxy.}
{The spectrometer SPI aboard INTEGRAL has accumulated over five
years of data on \Al gamma-ray emission from the Galactic plane.
We analyzed these data using suitable instrumental-background
models and adopted sky distribution models to produce
high-resolution \Al spectra of Galactic emission, spatially
resolved along the Galaxy plane.}
  {We detect the \Al line  from the inner Galaxy at
$\sim 28\sigma$ significance. The line appears narrow, and we
constrain broadening in the source regions to $<1.3$ keV
($2\sigma$). Different sky distribution models do not
significantly affect those large-scale results. The \Al intensity
for the inner Galaxy is derived as $(2.9\pm 0.2)\times 10^{-4}\
\mathrm{ph\ cm^{-2}\ s^{-1}\ rad^{-1}}$, consistent with earlier
results from COMPTEL and SPI data. This can be translated to an
\Al mass of $2.7\pm 0.7$ \ms in the Galaxy as a whole. The \Al
intensity is also confirmed to be somewhat brighter in the 4th
than in the 1st quadrant (ratio $\sim 1.3\pm 0.2$). \Al spectra
separately derived for regions along the Galactic plane show clear
line centroid shifts, attributed largely to the Galaxy's
large-scale rotation. The \Al line toward the direction of the
Aquila region ($20\degr < l < 40\degr$) appears somewhat
broadened. Latitudinal variations of \Al emission towards the
inner Galaxy are studied, finding a latitudinal scale height of
$130^{+120}_{-70}$~pc $(1\sigma)$ for \Al in the inner Galaxy and
a hint (3$\sigma$) of peculiar \Al emission towards the region
$l<0^\circ,\ b>5^\circ$. }
  {}

  \maketitle

\section{Introduction}

The unstable isotope \Al has a mean lifetime of 1.04 Myr. \Al
first decays into an excited state of $^{26}$Mg, which de-excites
into the $^{26}$Mg ground state by emitting gamma-ray photons with
the characteristic energy of 1809 keV . The 1809 keV gamma-ray
line from radioactive \Al serves as a tracer of the recent
nucleosynthesis activity in the Galaxy.

The 1809 keV $\gamma$-ray line emission from the Galaxy was first
detected with the Ge spectrometer on the HEAO-C space experiment
(Mahoney et al. 1982). The Compton Observatory sky survey
1991-2000 then with COMPTEL imaging of the \Al line across the sky
showed that \Al emission extends along the Galactic plane, thus
clearly establishing \Al nucleosynthesis as a widely-distributed
Galactic phenomenon (Diehl et al. 1995, Pl\"uschke et al. 2001).
The structure of this emission, alignments of emission maxima with
spiral-arm tangent, and comparisons with tracers of candidate \Al
sources, all point to a predominant origin of \Al in massive stars
(Prantzos and Diehl 1996, Chen et al. 1995, Diehl et al. 1995,
Kn\"odlseder 1999).

The detailed study of \Al line emission from the Galaxy is one of
the main science goals of the INTEGRAL mission. SPI aboard
INTEGRAL is a high-resolution spectrometer with energy resolution
of 3 keV (FWHM) at 1809 keV, which therefore adds high-resolution
spectroscopic information to \Al astronomy. The Compton
Observatory in its early design phase included a fifth instrument
`GRSE', a high-resolution spectrometer later abandoned for cost
and complexity reasons. COMPTEL as a scintillation-detector based
instrument had modest spectral resolution of about 10\% (FWHM).
The detailed measurement of \Al line position and shape is
expected to reveal more information beyond the COMPTEL imaging
survey about the \Al sources and their location through the
Doppler effect, induced from Galactic rotation and dynamics of the
ejected \Al as it propagates in the interstellar medium around its
stellar sources.

Earlier INTEGRAL analysis had used 1.5 years of SPI data to first
explore the large-scale spectral characteristics of \Al emission
in the inner Galaxy (Diehl et al. 2006a, 2006b). A detection of
the \Al line from the inner Galaxy with a significance of $\sim
16\sigma$ confirmed the narrowness of the \Al line (FWHM $<2.8$
keV, $2\sigma$), which had already been seen by RHESSI (Smith et
al. 2003) and HEAO-C (Mahoney et al. 1984) earlier. We use
INTEGRAL/SPI data accumulated from five years to extend this study
towards spatially-resolved details of \Al line spectroscopy across
the inner regions of the Galaxy. In this paper, we concentrate on
the spectral and intensity variations of \Al emission along the
Galactic plane. The main goal is to explore if there are
variations from the large-scale properties of \Al emission, such
as bulk motion or enhanced turbulence in specific regions. Our
study will be limited by the brightness of the signal per region,
and hence proceed from broad to more constrained regions along the
plane, as the current exposure and \Al brightness allows; with
more exposure, in particular towards outer regions of the Galaxy
at higher longitudes, the extended INTEGRAL mission is expected to
eventually allow such studies up to the spatial resolution of
about 2.8$^{\circ}$ (FWHM) of the SPI instrument. We also derive
spectra for separate regions of Galactic latitudes, to probe the
symmetry and the scale height of Galactic \Al emission.

In this paper, we first describe the SPI observations and
methods of data analysis. Then we proceed with an update of the
large-scale characteristics of Galactic \Al emission, before
refining the spatial resolution of our study. We discuss the
implications of our findings with respect to the large-scale
distribution of \Al sources, the interstellar medium in their
vicinity, and possible regional deviations.
Studies of specific regions and their \Al emission are underway
and will be reported in separate papers
 (Cygnus -- Martin et al., in preparation; Orion -- Lang et al.,
 in preparation; Sco-Cen -- Diehl et al., in preparation).

\begin{figure}
\centering
\includegraphics[angle=0,width=8.5cm]{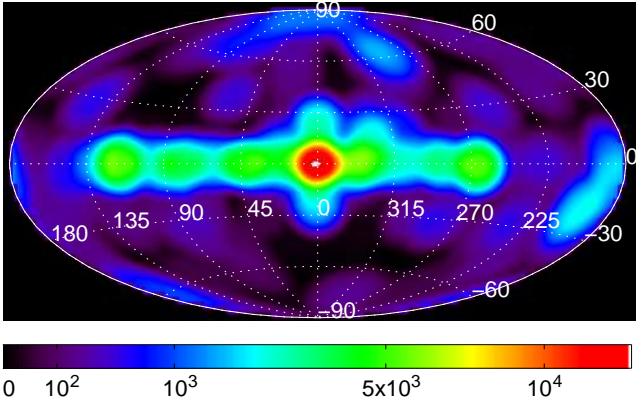}
\caption{Exposure of the sky in Galactic coordinates (the number
at the color bar in units of ksec) for the data selected from
5-year SPI observations for our \Al study (INTEGRAL orbits 43 --
650). The database covers the whole sky, with 29736 individual
pointings, equivalent to a total (deadtime-corrected) exposure
time of 61 Ms. Observations emphasize the Galactic plane,
specially in the inner Galaxy region, but also specific regions of
interest such as Cas A, Cygnus, Carina-Vela, Orion, Virgo, and the
Crab pulsar and nebula.}
\end{figure}

\section{Observations and Data Analysis}

\subsection{SPI observations}

The INTEGRAL spacecraft was launched on October 17, 2002, into a
high-inclination, high-eccentricity orbit intended to avoid the
increased background from the Earth's trapped radiation belts.
INTEGRAL's orbital period is $\sim$ 3 days. The spectrometer SPI
consists of 19 Ge detectors actively shielded by a BGO
anti-coincidence shield. It has a tungsten coded mask in its
aperture which allows imaging at $\sim 2.8^{\circ}$ resolution
within a $16^{\circ} \times 16^{\circ}$ full coded field of view
(imaging on INTEGRAL is mainly performed at lower energies by the
IBIS telescope, with which SPI is co-aligned). The Ge detectors
are sensitive to gamma-rays between 15 keV and 8 MeV, with a total
effective area $\sim 70$ cm$^{2}$ at 1 MeV, and achieve an energy
resolution of $\sim 2.5$ keV at 1 MeV (Roques et al. 2003,
Atti\'{e} et al. 2003). However, cosmic-ray (CR) impacts degrade
this resolution over time, and the instrument SPI is periodically
switched off for 14 days twice a year while annealing (by heating
from cryogenic temperatures $\sim$ 80 K to 100$^{\circ}$C) is
applied to the detectors to restore the energy resolution back to
its pre-launch value by thermal curing heating of the CR-induced
defects (Roques et al. 2003, Leleux et al. 2003).

In space operations, INTEGRAL with its IBIS and SPI telescopes is
pointed with a fixed attitude for intervals of typically $\sim
2000$ s (referred to as {\em pointings\/}), which are successively
arranged as a standard pattern of neighbouring pointings $\simeq
2^\circ$ apart ({\em dithering\/}), and covering target region of
interest for improved imaging (Jensen et al. 2003; Courvoisier et
al. 2003).

We obtain a database from the 5-year SPI observations, i.e. from
the INTEGRAL orbits 43 -- 650, which encompasses 29736 pointings
of the spacecraft and its instruments across the sky, equivalent
to a total deadtime-corrected exposure time of 61 Ms (see the
exposure map in Figure 1).

The instrument operation has been interrupted from a few short
anomalies and the occurrences of solar flares, and most
significantly for the typically two-week annealing episodes
described above, and by the regular perigee passages with
switch-offs due to the high background intensity from the
radiation belts. The sensitivity of our observations were further
reduced by the failure of two of the 19 detectors (December 2003
and July 2004).

\begin{figure}
\centering
\includegraphics[width=9.0cm]{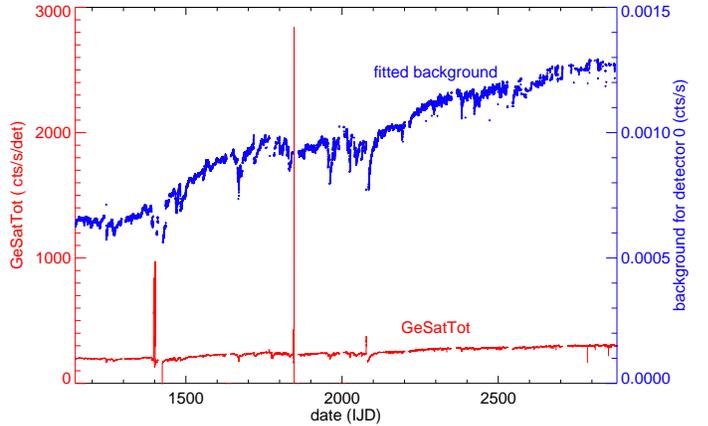}
\caption{The unfiltered total GEDSAT rates and background rates
for detector 0 around 1809 keV in model fittings versus time (the
INTEGRAL Julian Date, starts at 1 Jan 2000). The total GEDSAT
rates per detector generally vary in a range from 150 -- 300
counts/s, but some erratic cases like zero values and solar flare
spikes are excluded from science analysis with our primary
selections (also see the text). }
\end{figure}

\begin{figure*}
\centering
\includegraphics[width=8.5cm]{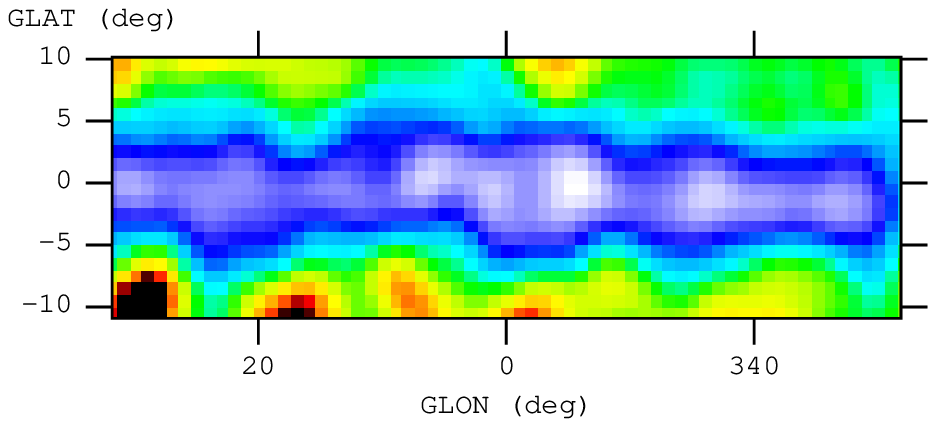}
\includegraphics[width=8.5cm]{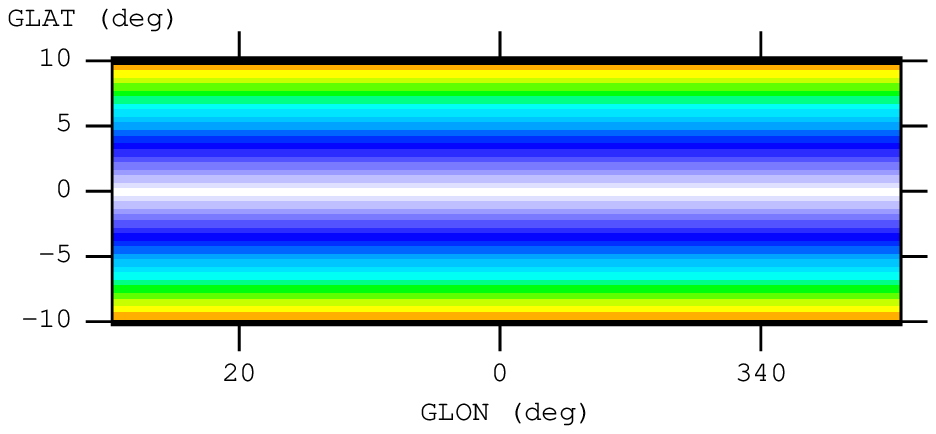}
\caption{The COMPTEL maximum entropy \Al map (left, Pl\"uschke et
al. 2001, hereafter COMPTEL MaxEnt) and a homogenous disk model
(right, hereafter Homo Disk, constant brightness along longitudes,
exponential-like in latitudes with scale height 200 pc) for the
inner Galaxy as the input sky models in the model fitting.  }
\end{figure*}

\begin{figure*}
\centering
\includegraphics[width=8cm]{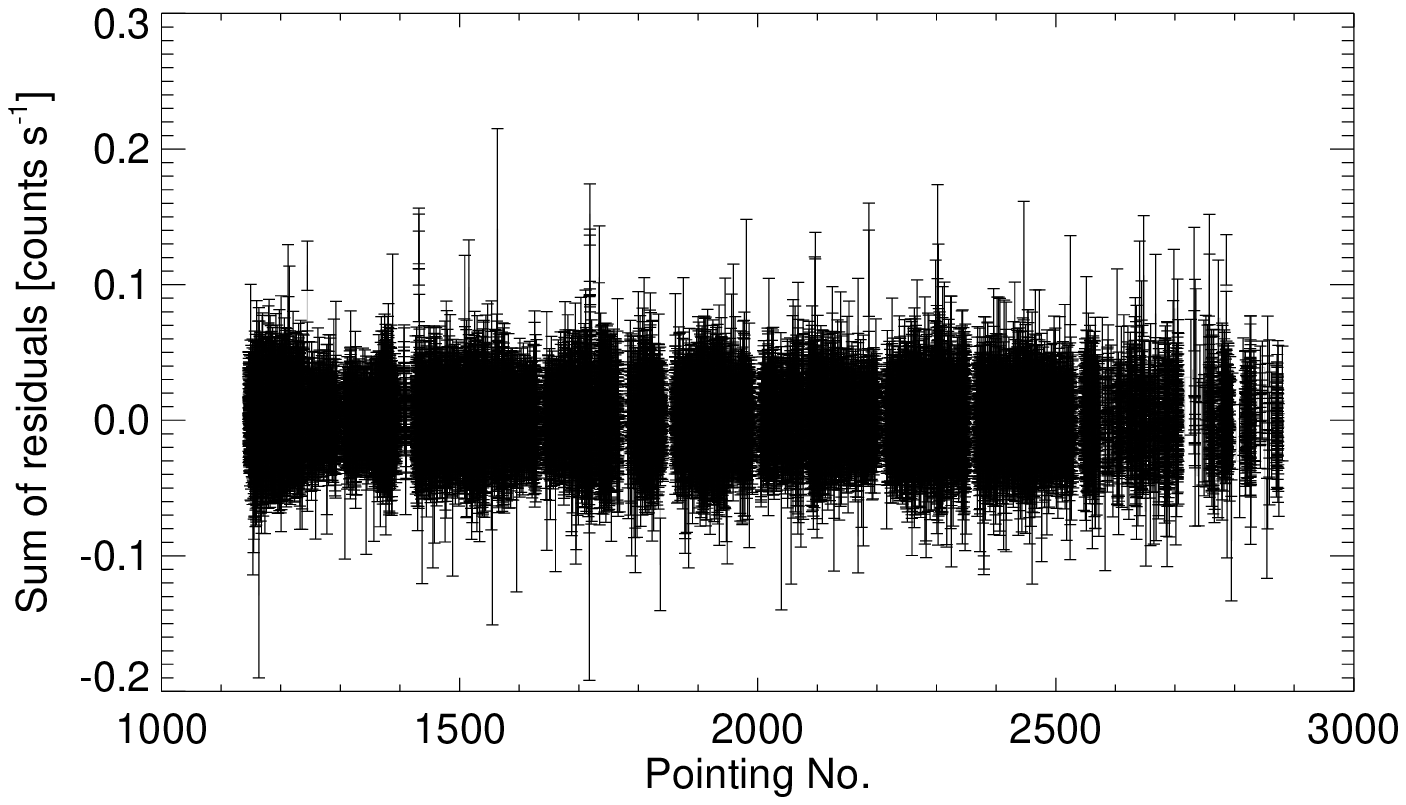}
\includegraphics[width=8cm]{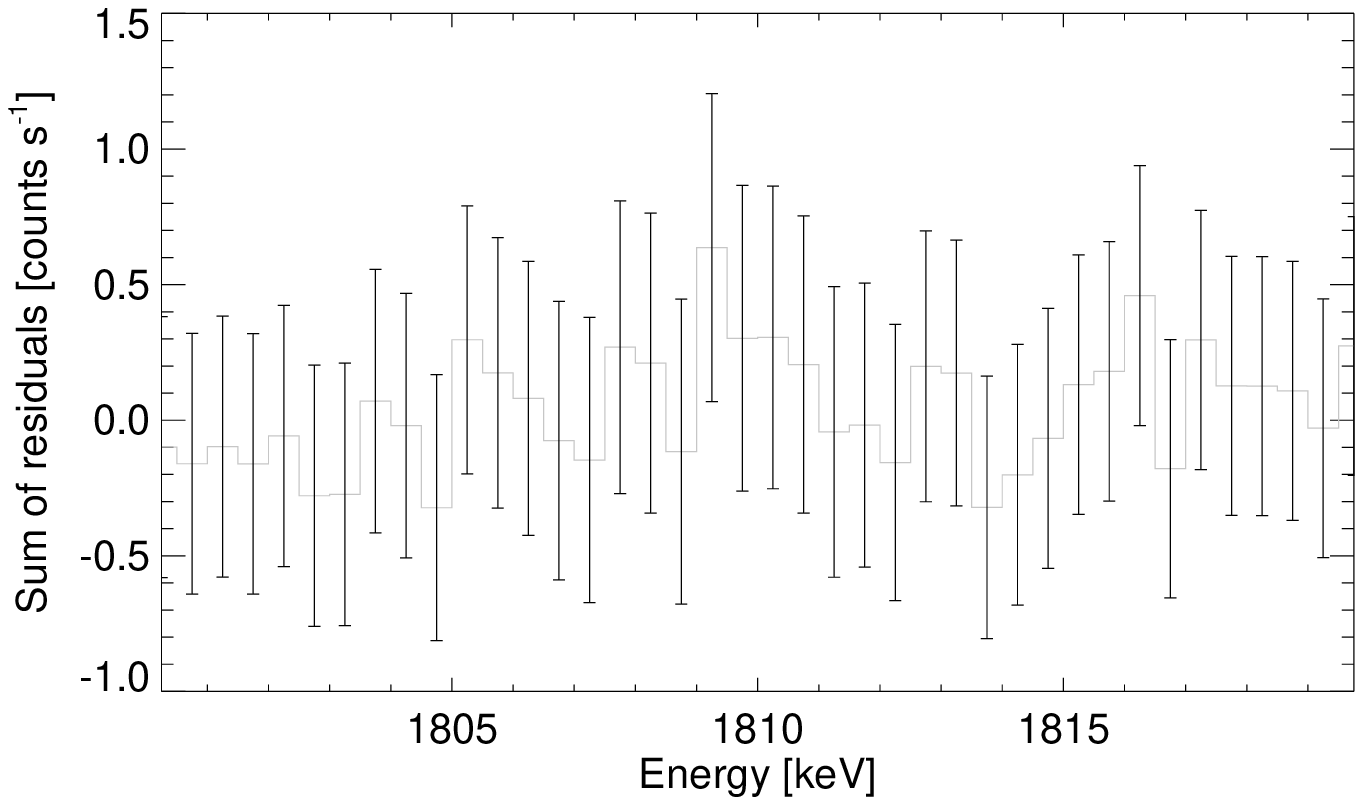}
\caption{Residuals versus time (pointings, per 30 min, left) and
energies (right) after model fitting of the inner Galaxy region
for the SE database. Residuals around zero confirm that our
background models are adequate, with $\chi^2/d.o.f. = 0.958$ in
model fittings.}
\end{figure*}

\subsection{Data Analysis}

Our data analysis steps include: (1) selection and assembly of
data which are free of contaminations by anomalies and, e.g.,
solar-flare events; (2) modelling the instrumental background; (3)
deriving spectra by fitting the measured data in narrow energy
bins with our background model and a model of the spatial
distribution of celestial gamma-ray emission, folded through the
instrumental response into the data space of the measurement.
These steps have been described in several papers, e.g. Strong et
al. (2005), Diehl et al. (2003), and as applied to gamma-ray lines
specifically in Wang et al. (2007) and Diehl et al. (2006b). Here
we give a concise summary and add comments as relevant for the
studies of this paper.

We first apply a selection filter to reject corrupted or invalid
data. This is done for data sections called `science window',
typically a time interval of $\sim 30$ minutes corresponding to
one pointing. We apply selection windows to `science housekeeping'
parameters such as the count rates in several background tracers,
proper instrument status codes, and orbit phase. For tracing
background, we employ the SPI plastic scintillator anticoincidence
counter (PSAC), and the rate of saturating events in SPI's Ge
detectors (i.e., events depositing $>$ 8 MeV in a single Ge
detector; hereafter referred to as GEDSAT rates, see Figure 2).
This leads to exclusions of solar-flare periods and other periods
of clearly increased / abnormal backgrounds. Additionally, regular
background increases during and after passages through the Earth's
radiation belts are eliminated through a 0.05--0.916 window on
orbital phase.

From the selected events, spectra are accumulated per each
detector and pointing in energy bins of 0.5~keV in the spectral
range of 1785 -- 1826 keV, and together with dead time and
pointing information assembled into the analysis database. From
identically-selected data, we also establish a database for the
background modelling from the adjacent continuum, combining
broader spectral ranges on both sides of the \Al line into single
bins.

We select events which trigger one and only one of our 19 Ge
detectors ('single events', SE). We exclude the $\simeq$50\%
fraction of 'multiple' events, because for events interacting in
more than one detector different combinations of single-detector
energies with their correspondingly-different resolutions would be
superimposed for identical values of total energy, thus leading to
an ill-determined spectral energy response. This reduces the total
signal obtained, but guarantees that the spectral response of the
instrument is well-defined for our dataset.

Instrumental background is dominated by a rather smooth spectral
continuum, with instrumental lines superimposed - for our case, a
broad instrumental-line feature centered at 1810~keV comprises
~30\% of the count rate (see Fig. 2 in Diehl et al. 2006b). For
our model of the continuum background, we make use of the
simultaneously-measured events in energy bands adjacent to the \Al
line region, at 1785~--~1802~keV plus 1815~--~1826~keV, and use
each detector's count rates in this adjacent range to normalize
the GEDSAT time series with their superior statistical precision
at regular intervals of several days. In Figure 2 we show both the
original GEDSAT count rate history and the background model for
one of our detectors, as it varies over the dataset. It is seen
that anomalous spikes of our background are excluded from our
data, and that there still is significant temporal structure in
our fitted background model. This `adjacent-energies' background
variation model per detector thus is based on high-statistics
GEDSAT rates over short time scales of $\sim$ 100 pointings, as
these have been found to trace the prompt CR activation rather
well. Adjusting these 'templates' to the actual counts per
detector in the 0.5 keV bins used for spectral analysis of the sky
signal then ensures that over longer time scales (typically 3
days, where counts in those narrow bins are statistically
sufficient) any second-order deviations from GEDSAT tracing actual
backgrounds are accounted for. Through this procedure we ensure
that pointing-to-pointing variations are given by our background
tracer, independent of the coded-mask orientation on the sky. The
broad instrumental line feature centered at 1810 keV is actually
found to also be rather well modelled by these 0.5~keV bin
adjustments, indicating that the line and continuum backgrounds
have similar temporal behaviour.

The cosmic-signal spectra are obtained from our measured database
of detector spectra when we combine above background models with a
spatial model for sky emission (Figure 3), allowing for
adjustments of intensity parameters for background and sky
intensities per energy bin:

\beq D_{e,d,p}=\sum_{m,n}\sum_{j=1}^{k_1}
A_{e,d,p}^{j,m,n}\beta_s^j I_j^{m,n} + \sum_t
\sum_{i=1}^{k_2}\beta^i_{b,t} B^i_{e,d,p} + \delta_{e,d,p}, \enq
where {\it e,d,p} are indices for data space dimensions: energy,
detector, pointing; {\it m,n} indices for the sky dimensions
(galactic longitude, latitude); $A$ is the instrument response
matrix, $I$ is the intensity per pixel on the sky,
$k_1$ is the number of independent sky intensity distribution
maps; $k_2$ is the number of background components, $\delta$ is
the count residue after the fitting. The coefficients $\beta_s$
for the sky map intensity are constant in time, while
$\beta_{b,t}$ is allowed time dependent, to cater for different
background normalizations for each camera configuration of
19/18/17 functional detector elements. The sky brightness
amplitudes $\beta_s$ comprise the resultant spectra of the signal
from the sky. We generally use a maximum-likelihood fitting method
(implemented in a SPI standard tool called {\em spimodfit},
properly accounting for Poisson statistics in our spectra; for
more details, see Strong et al. 2005).

We thus obtain per energy bin the fitted intensity parameter
values with their uncertainties, their covariance matrices, and
the fit residuals. Analysis of residuals, uncertainties, and
covariance matrices are made to determine the validity of the fit
of such a combined model to our measured set of spectra. Residuals
after the model fitting are shown in Figure 4 (residuals with the
time and energies). Reduced $\chi^2$ values around 1.0 confirm
that both our background model and the sky distribution model(s)
are adequate.

We use the sky intensity distribution of $^{26}$Al as derived from
9 years of COMPTEL observations (maximum entropy map, Pl\"uschke
et al. 2001) as the standard reference sky model. We compare this
to different \Al emission tracer models or maps, to assess the
impact of different sky distribution models on the \Al line shape
and intensity (see \S 3). For simultaneous fits of sets of
partial-sky distribution models (in order to derive
spatially-resolved spectra), we use analytical forms of sky
distribution models with prescribed symmetry to avoid biases to
such analysis (see below).

After we obtain fitted-amplitude coefficient spectra for the sky
components, we characterize these spectra and in particular the
\Al line details therein through two different approaches. When
the sky signal is weak, we fit the spectra with Gaussians plus a
linear residual background, and use the Gaussian intensities,
centroid energies, and FWHM widths for relative comparisons, such
as trends along the plane of the Galaxy. When the signal is
sufficiently strong so that we are sensitive to line width
details, we describe the line component not by a single Gaussian
any more, but rather by the convolution of a Gaussian with the
asymmetric instrumental line response as it develops from
degradation of our detectors' resolution and their periodic
restorations through annealings. This allows us to infer immediate
information about the celestial \Al dynamics, which we identify
with this Gaussian, and in particular its width, which arises from
Doppler shifting of the line energies with motion of decaying \Al
nuclei relative to the observer. In both cases, we fit these
spectral models to the amplitude values and their uncertainties as
determined from the model fit to the large set of spectra,
performing a maximum-likelihood fit with the Levenberg-Marquardt
algorithm. In the latter case, we derive the entire probability
distribution of the fitted parameter values through a Monte Carlo
Markov Chain (MCMC) method, which allows us in particular to
determine probability constraints for the celestial \Al line
broadening from kinematics. This parameter challenges the spectral
resolution of our instrument (about 3~keV at the \Al line energy),
hence so far is bounded from above only. Such asymmetric parameter
probability distributions may be far from Gaussian, hence require
such more sophisticated treatment (Kretschmer et al. 2009, in
preparation).

\section{\Al in the Inner Galaxy -- An Update}

The spectral characteristics of \Al emission in the inner Galaxy
serve to study the current nucleosynthesis activity and the
properties of the interstellar medium near the \Al sources on a
large-scale averaged scale. We define the ``inner Galaxy'' as the
region $-30^\circ < l < 30^\circ, \ -10^\circ < b <10^\circ$), and
may use this as a representative region for this purpose, since it
coincides with the bright ridge of observed 1809~keV emission as
observed along the plane of the Galaxy. We specifically exclude
regions at higher longitudes such as Cygnus: This region in
particular has been recognized as special, in its supernova to
Wolf-Rayet ratio, for example (Kn\"odlseder et al. 2004); other
emission at large longitudes also may be deviant from the Galactic
average and over-emphasized because nearby. We therefore believe
that using the $-30^\circ < l < 30^\circ$ region will give us a
more representative picture of \Al source environments in the
Galaxy.

\begin{figure*}
\centering
\includegraphics[angle=0,width=8.5cm]{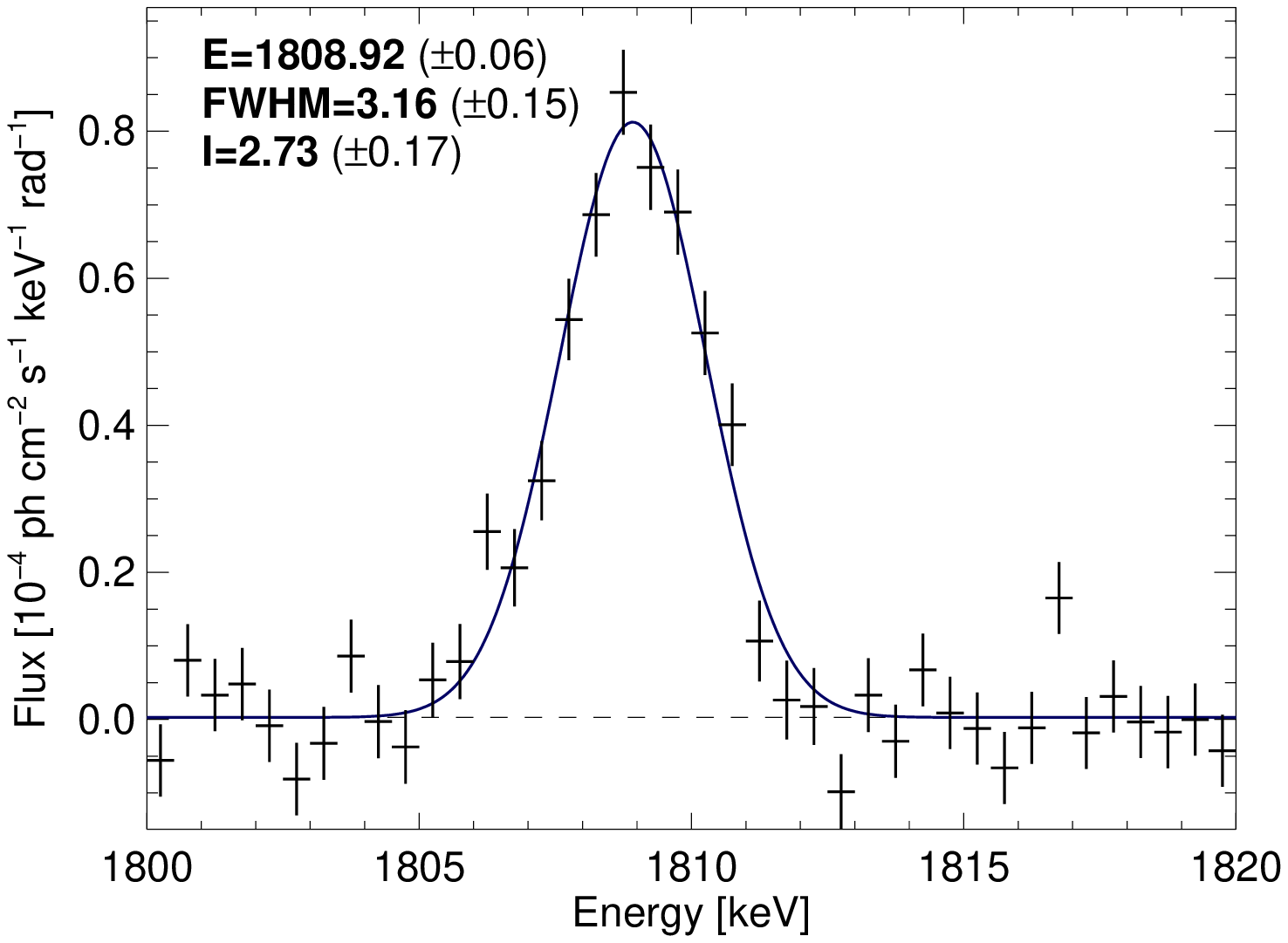}
\includegraphics[width=8.5cm]{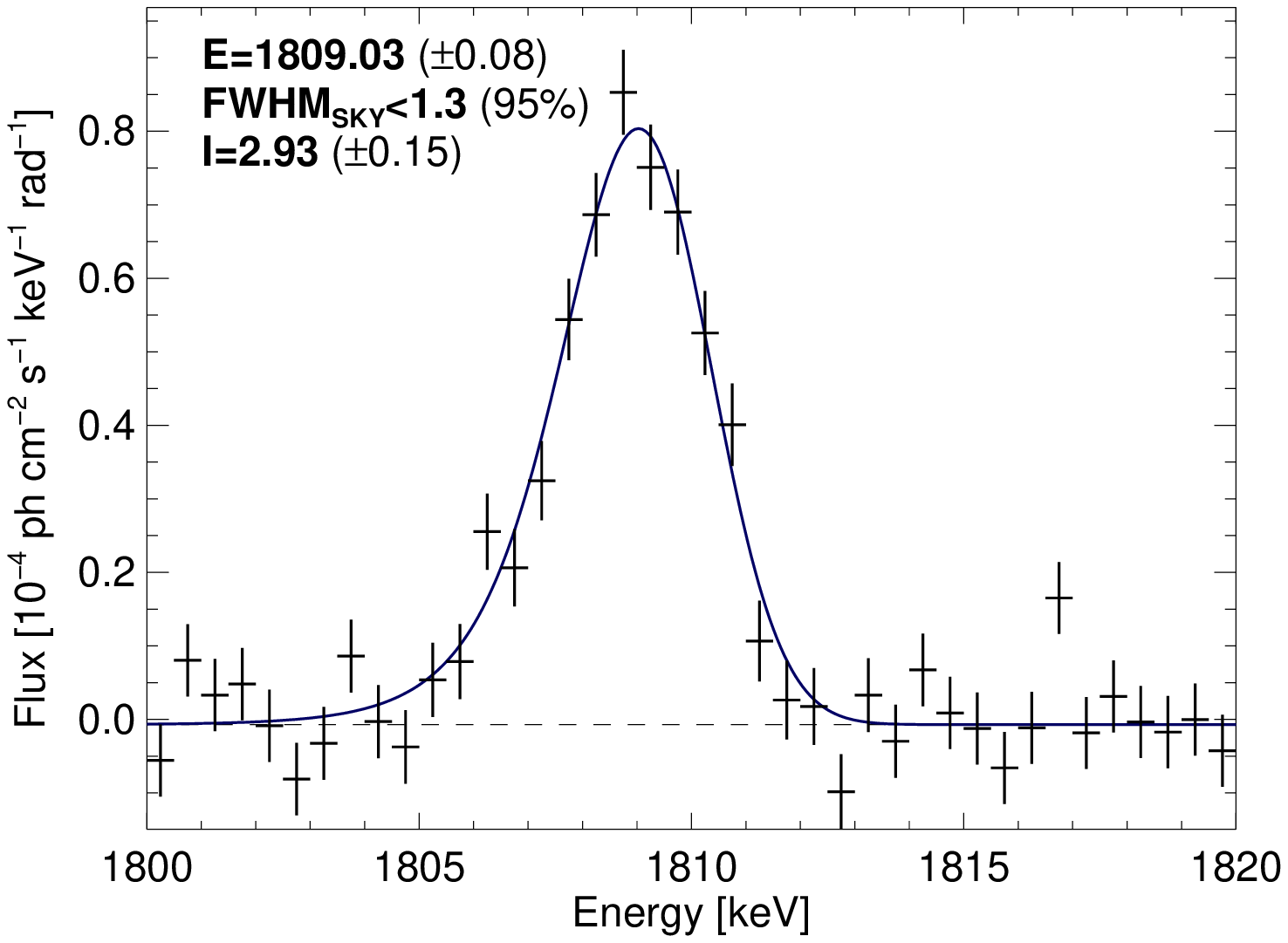}
\caption{Spectrum derived from sky model fitting using the COMPTEL
\Al Maximum Entropy image. The left figure shows the \Al line
fitted with a Gaussian, the width of $\sim$ 3.16 keV (FWHM) being
consistent with instrumental line widths around 1.8 MeV. The right
figure shows the line fitted with a composite line-shape model
using the time-averaged instrumental response as it results from
cosmic-ray degradation and annealings during the time of our
measurement, convolved with a Gaussian representing the cosmic
(intrinsic) \Al line width. The latter is found to be ($< 1.3$
keV, 2$\sigma$). Both fits find that the line is intrinsically
narrow. Systematic variations of derived line fluxes using two
spatial models are less than the statistical uncertainty in the
measurement (fluxes are quoted in units of $10^{-4}\ \mathrm{ph\
cm^{-2}\ s^{-1}\ rad^{-1}}$)}.
\end{figure*}

Fitting our set of observations to the sky intensity distribution
of the \Al maximum entropy image (MaxEnt) from COMPTEL (Figure 3,
left), we obtain the updated inner-Galaxy \Al emission spectrum
shown in Figure 5. MaxEnt is one possible sky intensity
distribution compatible with the COMPTEL data. The \Al line is
detected at $\sim 28\sigma$ significance.

The {\bf \Al gamma-ray flux} from the inner
Galaxy turns out as $(2.93\pm 0.15)\times 10^{-4}$ ph\ cm$^{-2}$\ s$^{-1}$\
rad$^{-1}$. This is consistent with our earlier values $(3.3\pm 0.4)\times 10^{-4}\
\mathrm{ph\ cm^{-2}\ s^{-1}\ rad^{-1}}$ (Diehl et al. 2006a,
2006b), and also with the COMPTEL imaging-analysis
value of $(2.8\pm 0.4)\times 10^{-4}$ ph\ cm$^{-2}$\ s$^{-1}$\
rad$^{-1}$ (Pl\"uschke et al. 2001).
The above value is derived using an asymmetric line shape as best matching
our expectations from SPI's spectral response and eventual additional celestial
line broadening the total \Al gamma-ray flux of
the Gaussian fit as determined for the inner Galaxy region is
$(2.73\pm 0.17)\times 10^{-4}\ \mathrm{ph\ cm^{-2}\ s^{-1}\
rad^{-1}}$.

In our model-fitting approach to derive \Al spectra from
SPI spectra per pointing (Eq. 1),  the adopted sky distribution models may
affect the \Al flux and line shape results.
We estimate the variations and potential systematic uncertainties
introduced by the spatial distribution model of the \Al emission
through variation of this sky distribution model within a plausible range.

Our ``standard'' is the measurement of \Al gamma-ray emission
directly, from COMPTEL. Uncertainties from COMPTEL imaging
analysis have however been shown to allow for a range of images
which are consistent with the COMPTEL measurements (Kn\"odlseder
et al. 1999). We consider the MaxEnt image the compromise between
making use of COMPTEL's imaging resolution of about 3.8$^\circ$
(FWHM) and the attempts to suppress artifacts from statistical
noise. A more conservative COMPTEL image is obtained from the
Multi-resolution Expectation Maximization method (MREM), which
carefully eliminates noise contributions at each iteration and
builds up the image starting from large spatial scales,
terminating once the image obtained does not need further
refinement in this statistical sense (Kn\"odlseder et al. 1999;
Kn\"odlseder 1999). The smooth MREM image from COMPTEL also shows
the inner Galactic ridge as well as Cygnus being bright in \Al
emission, yet does not show the few-degree scale features of the
MaxEnt map; we consider these two maps as adequate tests for
systematics from the range of direct \Al measurements.

Considering the limitations of gamma-ray telescopes, it has been
plausible to alternatively use maps obtained in
astronomically-more developed wavelength bands, once it it clear
that those trace \Al sources in the Galaxy. Detailed studies have
shown (Kn\"odlseder et al. 1999; Diehl et al. 1996) that among the
best tracers of \Al sources are (1) the infrared emission of warm
dust grains, mapped with the COBE/DIRBE and arising from radiative
heating of dust around clusters of massive stars (Bennett et al.
1996), and (2) radiation of free electrons (free-free emission,
Bremsstrahlung) observed at radio frequencies with the WMAP
satellite, arising from the ionizing massive-star radiation around
massive star clusters. From astrophysical arguments, also maps of
interstellar gas in different forms should trace the locations and
space density of \Al sources. (3) Molecular gas is observed
through CO line emission at radio frequencies, and has been mapped
in rather fine resolution (Dame et al., 1987 and 2001), (4) atomic
hydrogen (HI) sky surveys have been accumulated (e.g. Dickey and
Lockman 1990), and (5) cosmic-ray interactions with interstellar
gas produces penetrating continuum gamma-ray emission which has
been mapped in the EGRET sky survey (Hunter et al. 1997).

Finally, analytical models for the distribution of \Al sources
have been constructed, based on above knowledge of Galactic
structure in its different components (Robin et al. 2003),
properly weighted from astrophysical arguments. (6)
Double-exponential functions (in galactocentric radius, and scale
height above the Galactic plane) have been constructed, as well as
more sophisticated models including (7) spiral structure
components and building on the distribution of free electrons in
the Galaxy as derived from pulsar dispersion measurements (Taylor
\& Cordes 1993, and Cordes \& Lazio 2002). The scale height of \Al
sources has been found to lie between the molecular disk (about
50~pc) and the thick disk (about 0.3-1~kpc), with plausible values
around 200~pc.

We compare \Al line spectra determined from these different models
and tracers of \Al sources in the Galaxy. Fifteen different sky
distribution maps have been analyzed, and variations on \Al
brightness, \Al line centroid and width parameters are shown in
Figure 6. Systematic variations are within statistical
uncertainties, when we vary the spatial models for \Al emission.
We use their scatter to estimate a ``systematic'' uncertainty,
which turns out as $(2.9\pm 0.2)\times 10^{-4}{\rm ph\ cm^{-2}\
s^{-1}\ rad^{-1}}$, $1809.0\pm 0.09$ keV, $0.5 \pm 0.45$ keV for
the flux, centroid, and width parameters (with 1$\sigma$ error
bars), separately.

As before (Diehl et al., 2006a), we convert our measured \Al
intensity into an estimate of the {\bf total current \Al mass} in
the Galaxy, using an assumed geometrical source-distribution model
to extrapolate across the entire Galaxy from our inner-Galaxy
normalization of such a model, as discussed above. This is
required because the flux to mass conversion relies on \Al source
distances, not measured directly in projected sky brightness distribution
maps. Several three-dimensional distribution models are applied and
compared here, e.g.,  an exponential disk model, a ``young-disk''
model, a geometrical representation based on dust emission, and
a multi-component model including spiral-arm structures and based
on abundances of free electrons in interstellar space.
We vary scale height parameters of the appropriate components
to also study the latitude extent of the \Al emission. In all models,
we have taken the distance of the Sun to the Galactic center as
$R_0=$ 8.5~kpc (\Al mass sensitively depends on $R_0$, smaller
$R_0$ will globally reduce the size of the Galaxy, and result in
the smaller amount of \Al).

The average measured \Al flux of $(2.9\pm 0.2) \times 10^{-4}\
\mathrm{ph\ cm^{-2}\ s^{-1}\ rad^{-1}}$ for the inner Galaxy thus
translates into a Galactic \Al mass of $(2.7\pm 0.7)$ \ms using a
plausible scale height of 180 pc. If we ignored the ejection of
\Al into surrounding cavities and corresponding champagne flows
(see our own scale height determination below), and used the lower
scale heights of O stars or of the molecular disk, we would obtain
lower total amounts around or even below 2 \ms. This emphasizes
the need for spatially-resolved \Al studies (see Sect 5 and Fig.
12 below). The quoted uncertainty here includes both the
statistical uncertainty as propagated from the fitting method, and
the systematical uncertainty derived from variations among
plausible representations of the sky distribution of the emission
plus alternative 3-dimensional \Al source distribution models (see
also Figure A4.1 in Supplemental materials of Diehl et al. 2006a).

The {\bf line width} of \Al conveys global information about the
spread in projected velocities of \Al nuclei when they decay and
emit characteristic 1809~keV gamma-rays. Astrophysical origins of
\Al line width may come from two effects: random motions in the
interstellar medium (Chen et al. 1997) and Galactic differential
rotation (Kretschmer et al. 2003). A line broadening of $\sim$ 1
keV from our line-shape constraints corresponds to thermal Doppler
velocities of $\sim$ 120 km s$^{-1}$.

The Gaussian used to represent the \Al line (Fig.~5 left) shows a
width of $3.16\pm 0.15$ keV (FWHM), consistent with the
instrumental width determined as $\sim 3.1$ keV near 1809~keV from
nearby instrumental lines at 1764 and 1779~keV. This indicates
that the cumulative \Al line emission in the inner Galaxy is
intrinsically rather narrow in energy around the laboratory value,
with little kinematic Doppler broadening, well below SPI's energy
resolution.

\begin{figure*}
\centering
\includegraphics[width=5.5cm]{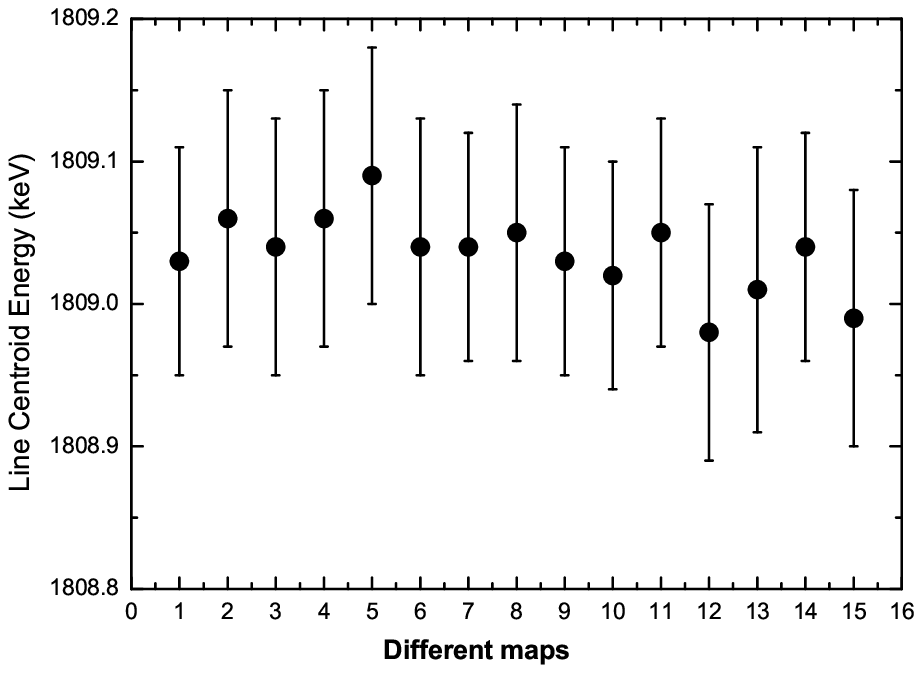}
\includegraphics[width=5.5cm]{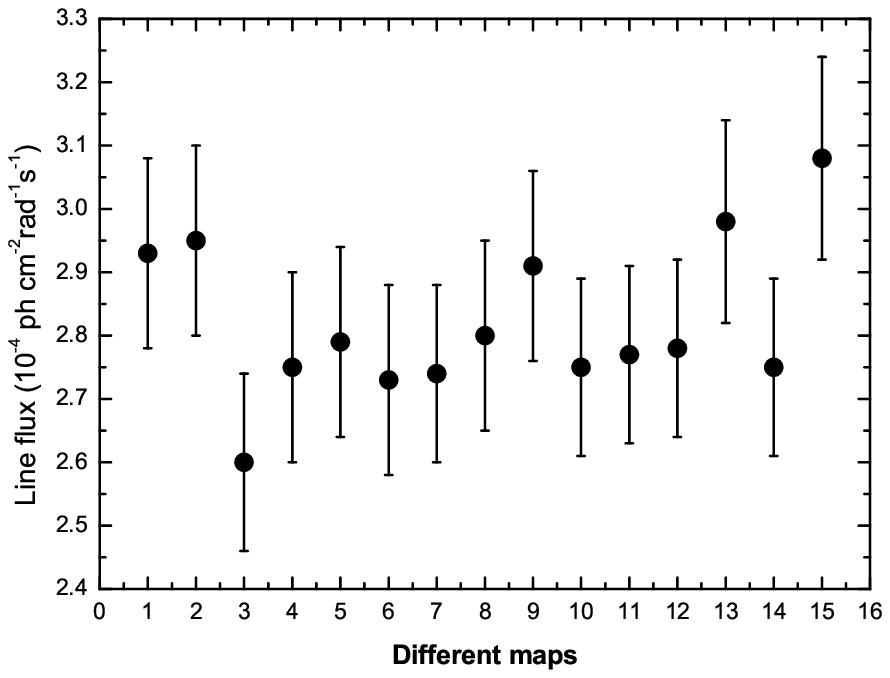}
\includegraphics[width=5.5cm]{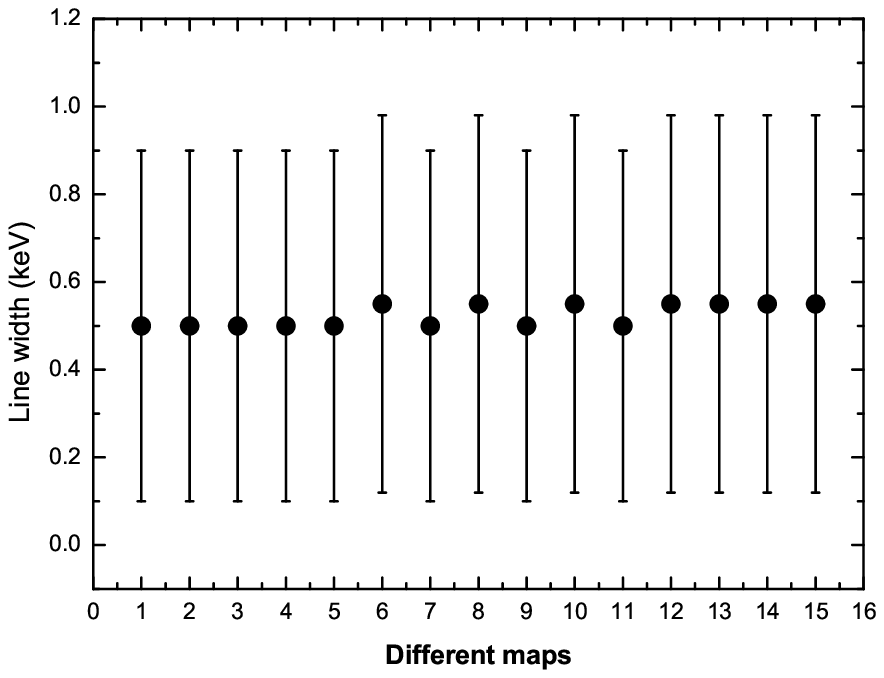}
\caption{Comparison of the three line parameters (centroid energy,
flux, width, with 1 $\sigma$ error bars) derived in model fittings
for different distribution models: 1. COMPTEL maximum entropy \Al
emission map (Pl\"uschke et al. 2001); 2. COMPTEL MREM \Al
emission map (Pl\"uschke et al. 2001); 3. an exponential disk
model (scale radius 4 kpc, scale height 180 pc); 4. a homogenous
disk model (Fig. 3); 5. HI (Dickey \& Lockman 1990); 6. CO (Dame
et al. 1987); 7. radio 408 MHz (Haslam et al. 1995); 8. DIRBE/COBE
240$\mu$m (Bennett et al. 1996); 9. IRAS 12$\mu$m (Wheelock et al.
1991); 10. EGRET ($>100 $ MeV, Hunter et al. 1997); 11. a young
disk model (Robin et al. 2003); 12. the free electron density
distribution model (TC93, scale height 150 pc, Taylor \& Cordes
1993) 13. TC93, scale height 300 pc; 14. the free electron density
distribution model (NE2001, scale height 140 pc, Cordes \& Lazio
2002); 15. NE2001, scale height 330 pc. }
\end{figure*}

For such a strong signal, we can be more ambitious, however, and
attempt to quantitatively constrain the intrinsic (celestial) line
broadening in a statistical probability analysis as described
above (see also Diehl et al. 2006b and Kretschmer et al. 2004). We
know the instrumental line response of our instrument at each
epoch of the mission from detailed studies of a large number of
instrumental lines. The response gradually degrades and develops
an increasing tail component on the low-energy side of the
photopeak energy, due to incomplete charge collection as the
detector's Ge lattice experiences damaging from cosmic-ray
impacts. Annealings restore better charge collection and a
symmetric instrumental-line response whose width is determined and
dominated by statistics of the charge collection. We assemble the
effective instrumental-line response for our study through
appropriate weightings of these time-dependent responses. Then we
fit the convolution of a parameterized Gaussian with this response
shape function, and thus derive values for the celestial \Al line
centroid energy and broadening. MCMC sampling of the probability
distributions of these parameters then allows us to translate the
parameter value uncertainties into probability constraints by
integrating over these distributions. In particular, the
probability distribution of the intrinsic width may not peak at
zero, indicating a small but by itself non-detectable broadening
of the line. In this case, our approach yields a reliable estimate
of the upper bound on line broadening, which is the
astrophysically-relevant quantity to constrain kinematics of
decaying \Al nuclei in the interstellar medium. The fitted
parameters are the line centroid, the intrinsic width of celestial
\Al, the intensity of the line, and two parameters for the
underlying continuum (see Figure 5 right). The intrinsic line
width is now constrained to $<1.3$ keV (2 $\sigma$, which is
consistent and yet significantly smaller than the earlier
constraints based on fewer SPI data (Diehl et al. 2003, 2006b).

\begin{figure*}
\centering
\includegraphics[width=8.5cm]{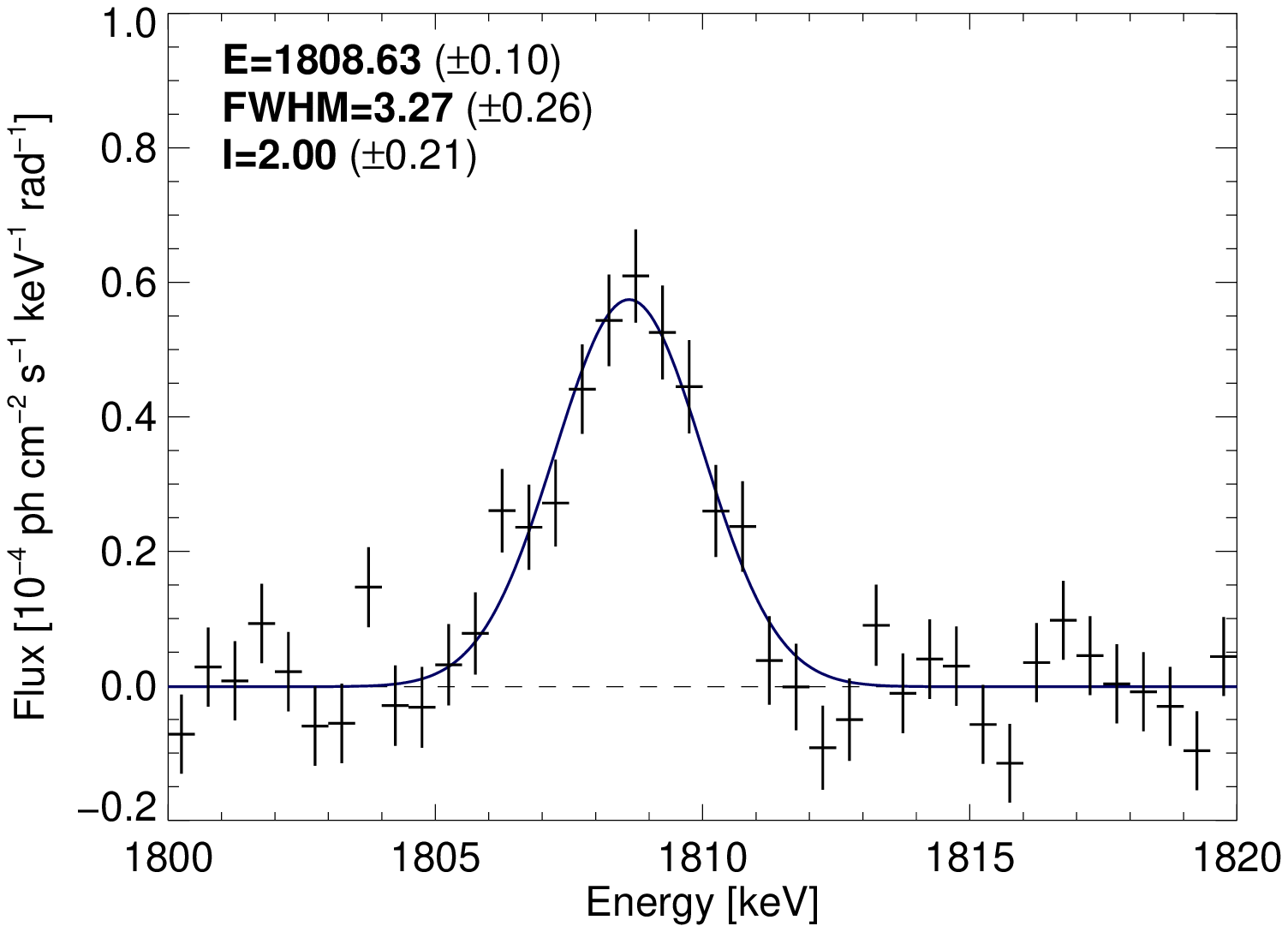}
\includegraphics[width=8.5cm]{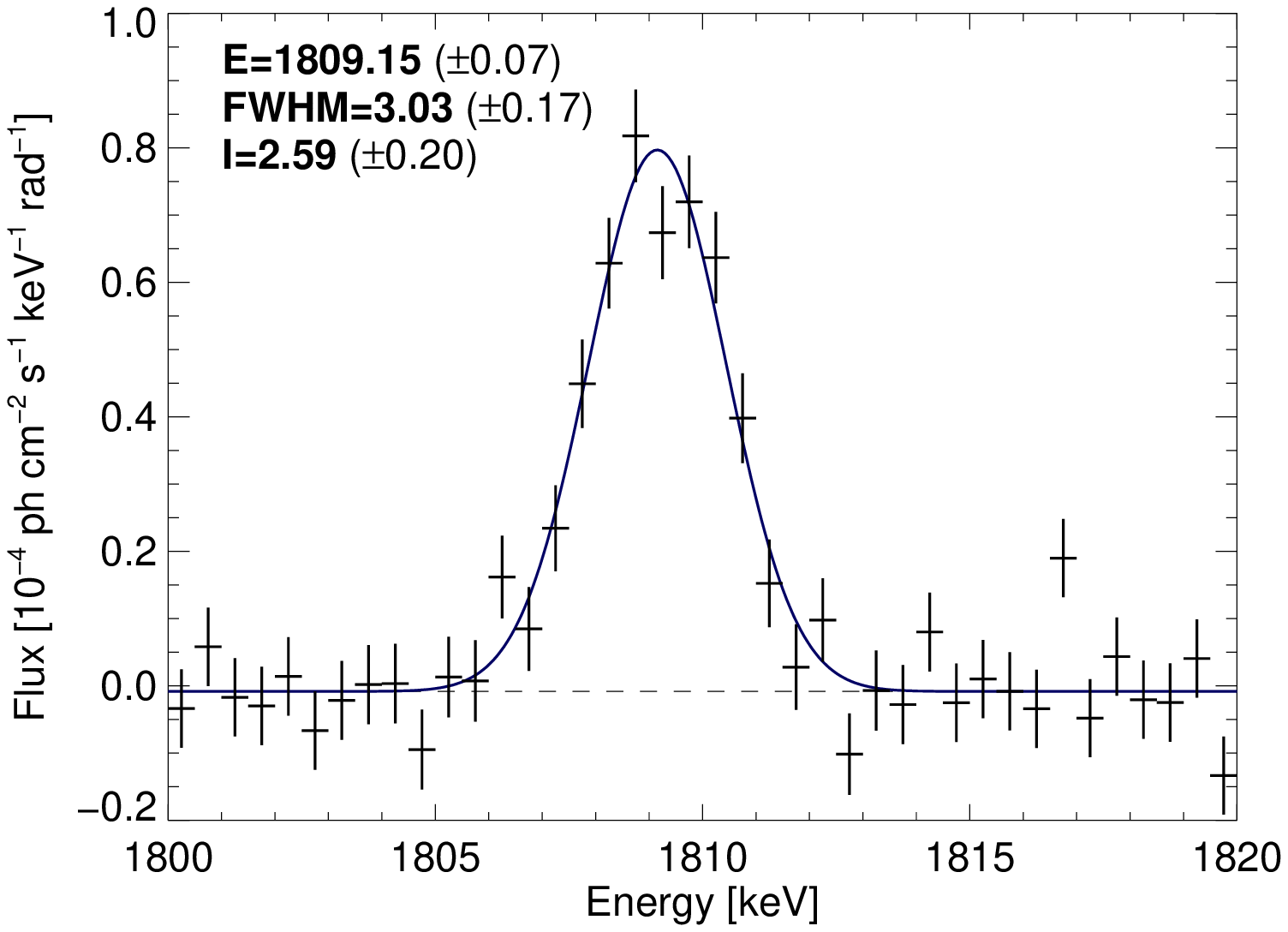}
\caption{\Al spectra for two Galactic quadrants ({\bf left}
$0^\circ < l < 60^\circ$, and {\bf right} $-60^\circ < l <
0^\circ$). Line centroids relative to the centroid energy of \Al
line in the laboratory (1808.65 keV) show a significant blueshift
in the 4th quadrant. Both the spectra have width values near the
value of the instrumental line width, implying that \Al emissions
from two quadrants are the narrow lines. In addition, the \Al flux
of the 4th quadrant is higher than that of the 1st quadrant, and
the flux ratio is $\sim 1.3$. }
\end{figure*}

In the inner Galaxy,  Galactic differential rotation alone can
lead to significant Doppler shifts towards specific longitudes
where the projected-velocity differences with respect to the solar
orbit reach maxima; line broadening results if we integrate over a
larger longitude range with different bulk velocity differences.
Kretschmer et al. (2003) have simulated the \Al line shape
diagnostics in the inner Galaxy due to Galactic rotation and \Al
ejection from sources, and find that line broadening of up to 1
keV is expected if the signal is integrated over the inner region
of the Galaxy. Our present large-scale line-shape constraints are
consistent with these expectations.

If we interpret line broadening of the \Al line from the inner
Galaxy in terms of interstellar-medium characteristics, the
intrinsic-width constraint of $<1.3$ keV corresponds to 160
km~s$^{-1}$ as a corresponding 2$\sigma $ limit on ISM velocities.
This is well within the plausible and acceptable range for the
environment of normal interstellar-medium turbulence (Chen et al.
1997), leading us to conclude that, within uncertainties, the
average velocities of decaying \Al in the Galaxy are not
abnormally-high (compare discussion after Naya et al. 1996 in Chen
et al. 1997).

In summary, the measured line width of \Al from the inner Galaxy
is consistent with Galactic rotation and modest
interstellar-medium turbulence around the sources of \Al. This
confirms earlier results on the \Al line width from HEAO-C
(Mahoney et al. 1984), RHESSI (Smith 2003), and SPI on INTEGRAL
(Diehl et al. 2006b). The GRIS balloon experiment had reported a
very broad line with a width $\sim 5.4$ keV (Naya 1996), which is
inconsistent with these measurements and clearly ruled out. This
is reassuring: within uncertainties, the average velocities of
decaying \Al in the Galaxy are not abnormally-high (see discussion
in Chen et al. 1997).

\begin{figure*}
\centering
\includegraphics[angle=0,width=7.5cm]{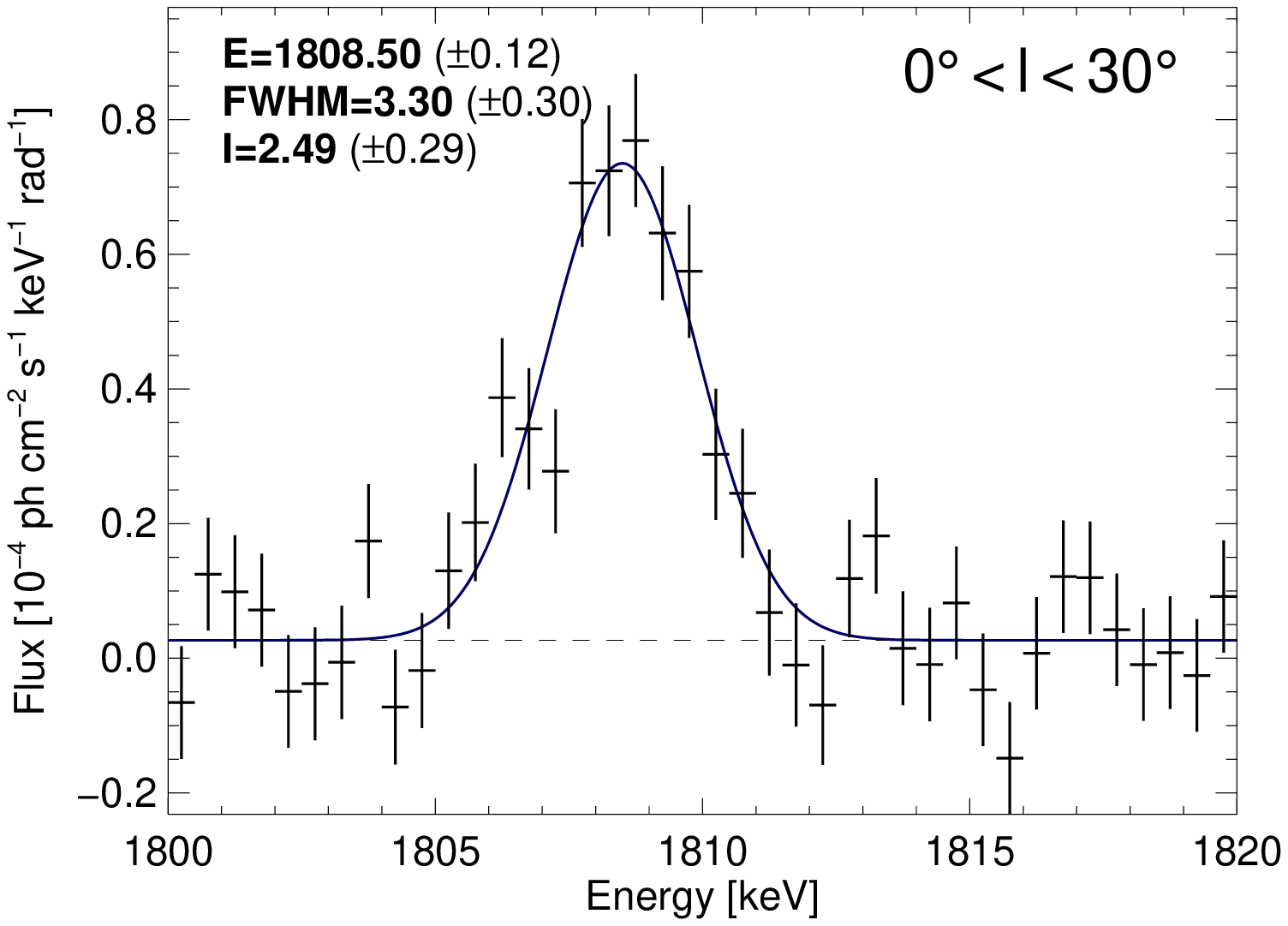}
\includegraphics[angle=0,width=7.5cm]{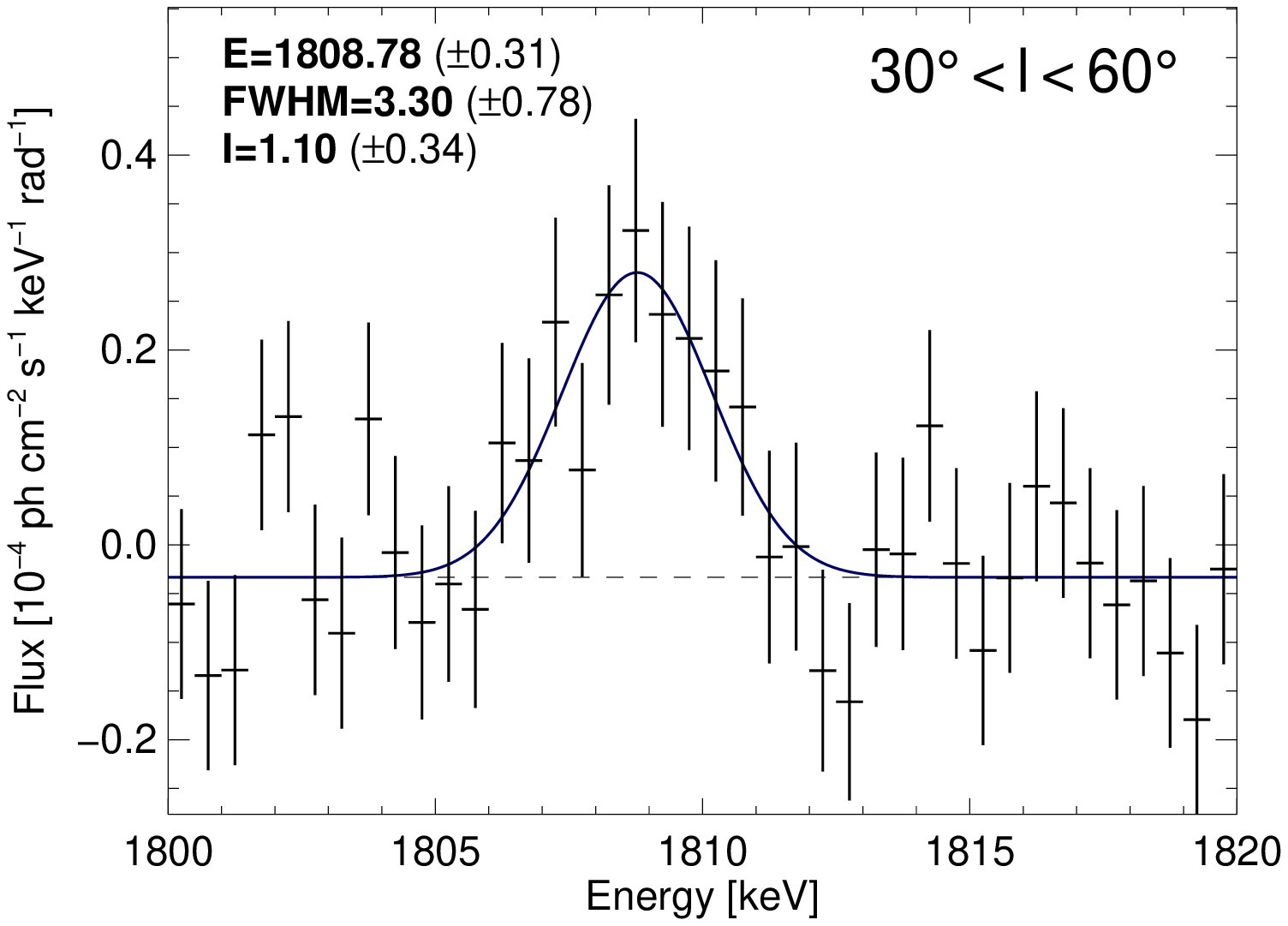}
\includegraphics[angle=0,width=7.5cm]{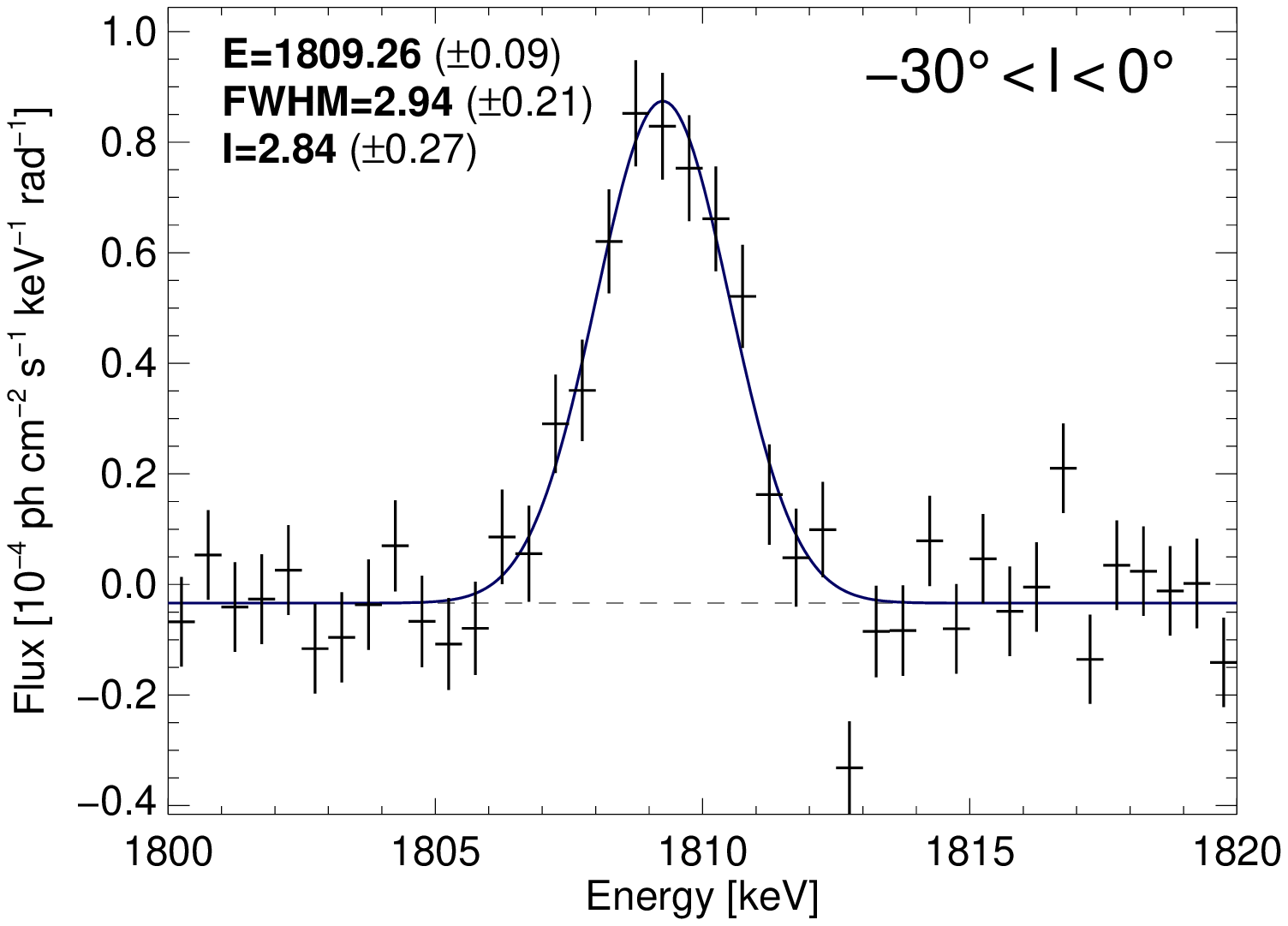}
\includegraphics[angle=0,width=7.5cm]{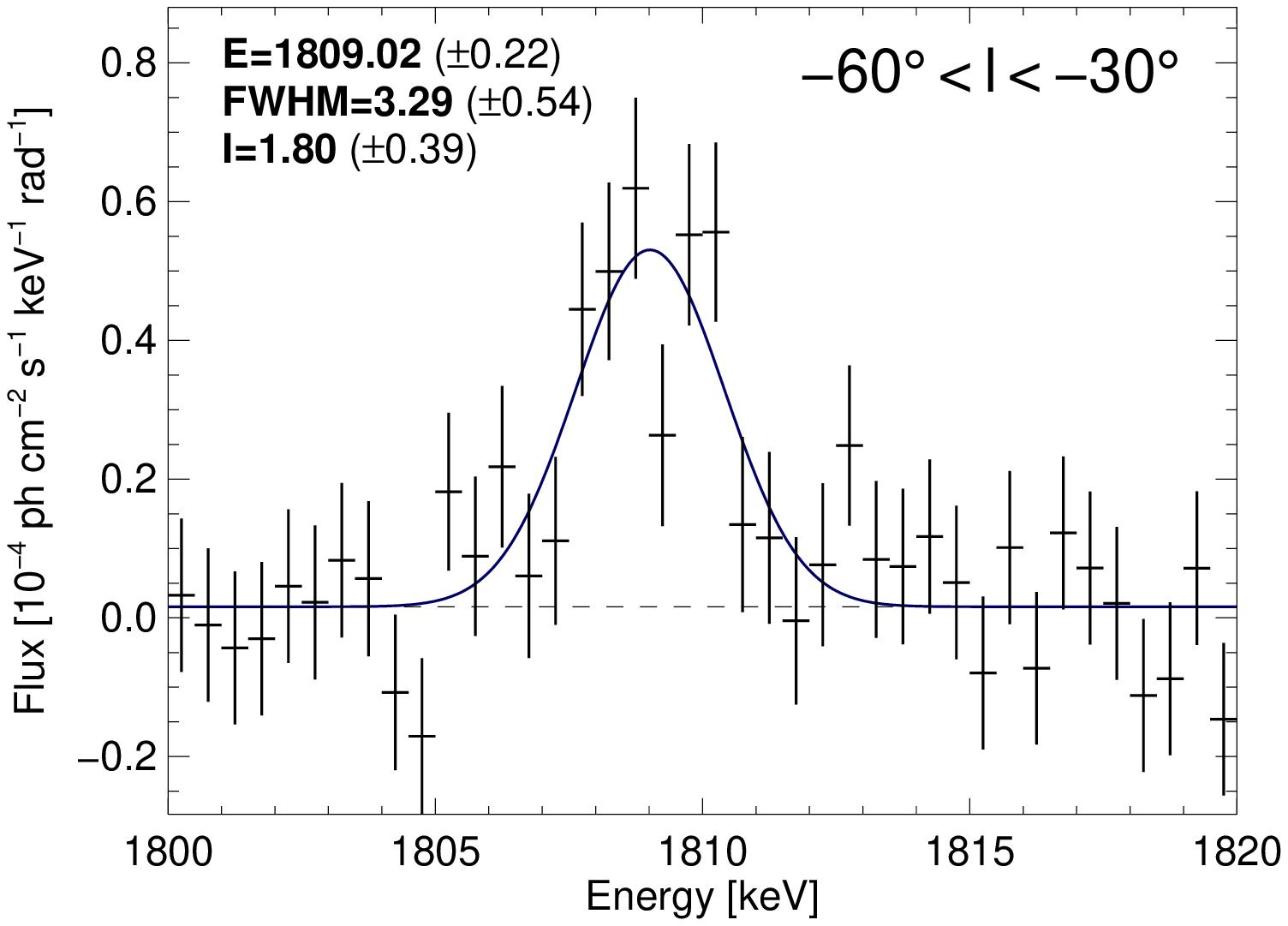}
\caption{\Al spectra of four segments along the Galactic plane (
$-60^\circ < l < 60^\circ$).  }
\end{figure*}

The relative position of {\bf the \Al gamma-ray line centroid}
with respect to the laboratory-determined \Al decay energy value
of 1808.65 $\pm 0.07$~keV (Firestone \& Ekstr\"om 2004) yields
information on potential bulk motion of \Al in the Galaxy. The \Al
line centroid energy from the whole inner Galaxy is determined at
$1808.92\pm 0.06$ keV from the Gaussian fit and $1809.03\pm 0.08$
keV from the instrumental-response-convolved fit, even after
considering uncertainties, the measured centroid energy is  higher
than the laboratory value. The effect of this blueshift for the
whole inner Galaxy will be discussed below.

In our determination of \Al line properties, we represent the
spectra derived from detailed model fitting by a line component
for \Al plus a linear component which could capture any underlying
systematics in our background or sky modellings. We expect that
our adjacent-energy background model component will eliminate
Galactic continuum emission, at least to first order, if the
spectral shape is rather flat across our 40~keV energy range
(1785--1826~keV). The diffuse gamma-ray continuum in the inner
Galaxy is $\sim 2\times 10^{-6}\ \mathrm{ph\ cm^{-2}\ s^{-1}
rad^{-1}\ keV^{-1}}$ in the energy band of 1 -- 2 MeV (Strong et
al. 1999). Indeed, we find that offsets above zero in our spectra
are rather small and negligible, supporting this property of our
background model.

\section{\Al emission along Galactic longitudes}

The 9-year COMPTEL imaging of \Al line emission had already
suggested some asymmetry in the inner Galaxy: the fourth Galactic
quadrant appears somewhat brighter than the first quadrant
(Pl\"uschke (2000) finds a significance of 2.5$\sigma$ for a
brightness difference). COMPTEL could provide the image details of
\Al in the Galaxy, but no significant spectral information due to
its spectral resolution of about 150~keV near the \Al line. SPI
with its Ge detectors features sufficiently-high spectral
resolution to allow astrophysical constraints from \Al line
shapes, averaged over the Galaxy as discussed above, but also for
different regions along the Galactic plane due to its imaging
properties as a coded-mask telescope.

\begin{figure}
\centering
\includegraphics[width=8.5cm]{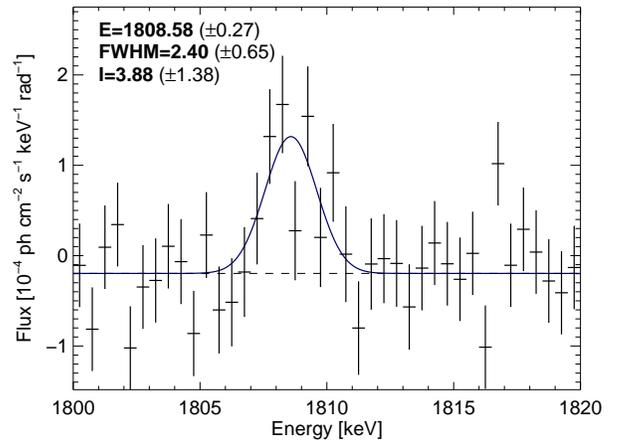}
\caption{\Al spectrum from the  Galactic center ($-5^\circ<l<
5^\circ,\ -10^\circ<b< 10^\circ$). The determined line centroid
energy ($1808.58\pm 0.27$ keV) is consistent with the laboratory
value.}
\end{figure}

In this section, we will proceed towards increasing spatial
resolution along the Galactic plane, starting out from testing
Galactic asymmetries between the first and fourth quadrant. \Al
line parameters toward the different directions of the Galactic
plane are determined using separate sky maps covering each sky
region, simultaneously fitting these together with our background
model to the entire sky survey database. \Al line fluxes, centroid
energies, and line widths then are derived by a simple Gaussian
fit to the \Al line in the resulting spectra, as we are interested
in relative changes between different portions of the sky. This
will allow us to identify line shifts from bulk motion such as
expected from large-scale Galactic rotation, and hints for
additional line broadenings in particular regions, which would
reflect increased \Al velocities in such regions. A homogenous
disk model (see Figure 2 right, $-60^\circ < l < 60^\circ, \
-10^\circ < b < 10^\circ$, scale height 200 pc ) is used here to
avoid a bias of sky distribution models along the Galactic plane,
for such relative comparison.

\Al spectra for the 1st ($0^\circ < l < 60^\circ$) and 4th
quadrant ($-60^\circ < l < 0^\circ$ ) are presented in Figure 7.
In the 4th quadrant we note a blueshift of $0.49\pm 0.07$ keV
relative to the centroid energy of \Al line in the laboratory, but
no significant redshift in the 1st quadrant is apparent ($\sim
0.04\pm 0.10$ keV). Both spectra have width values compatible with
no significant  \Al line broadenings. The indicated \Al asymmetry
between the two inner Galactic quadrants appears again, with a
flux ratio of $\sim 1.3\pm 0.2$.

We proceed further towards more confined Galactic regions using
four sub-maps with 30 degree width along Galactic longitude, to
obtain \Al spectra for these regions shown in Figure 8: (1)
$0^\circ < l < 30^\circ$, (2) $30^\circ < l < 60^\circ$, (3)
$-30^\circ < l < 0^\circ$, and (4) $-60^\circ < l < -30^\circ$.
Centroid energy shifts of the \Al line are found, $+0.15\pm 0.12$
keV and $-0.61\pm 0.09$ keV in regions (1) and (3) respectively,
as expected from large-scale Galactic rotation. The inner region
($-30^\circ < l < 30^\circ$) is seen to be much brighter than the
two outer regions of the Galaxy, consistent with the COMPTEL \Al
map (note that here we use a homogeneous sky distribution model as
the model-fitting prior). The indicated \Al emission asymmetry for
the 1st and 4th quadrants also shows up again between regions (3)
and (1), with a flux ratio of $\sim 1.15\pm 0.18$. Even further
out, region (4) may also be brighter than (2) by $\sim 1.6\pm
0.6$. Since smaller regions include less \Al signal, these
differences are, however, not significant.

Challenging the imaging capability of SPI for diffuse and extended
emission, we refine spatial structure even more towards smaller
longitude intervals. In Figure 9, we show a spectrum representing
the Galactic center region ($-5^\circ<l< 5^\circ, -10^\circ
<b<10^\circ$), where the \Al line is still significant ($>
4\sigma$). The line centroid energy is determined at $1808.58\pm
0.27$ keV, and is consistent with the laboratory value -- no line
shift from bulk motion is indicated.

\begin{figure*}
\centering
\includegraphics[angle=0,width=5.5cm]{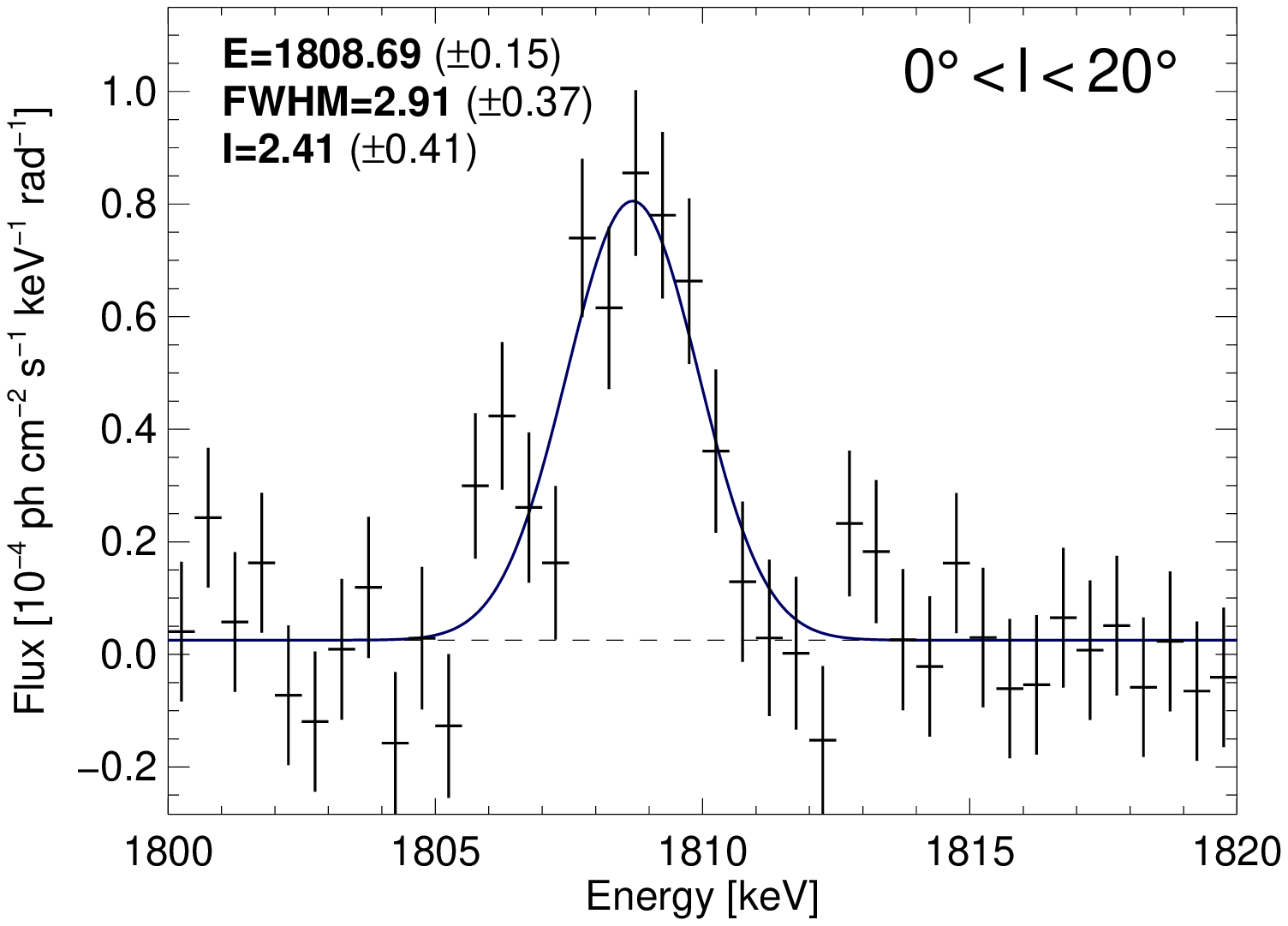}
\includegraphics[angle=0,width=5.5cm]{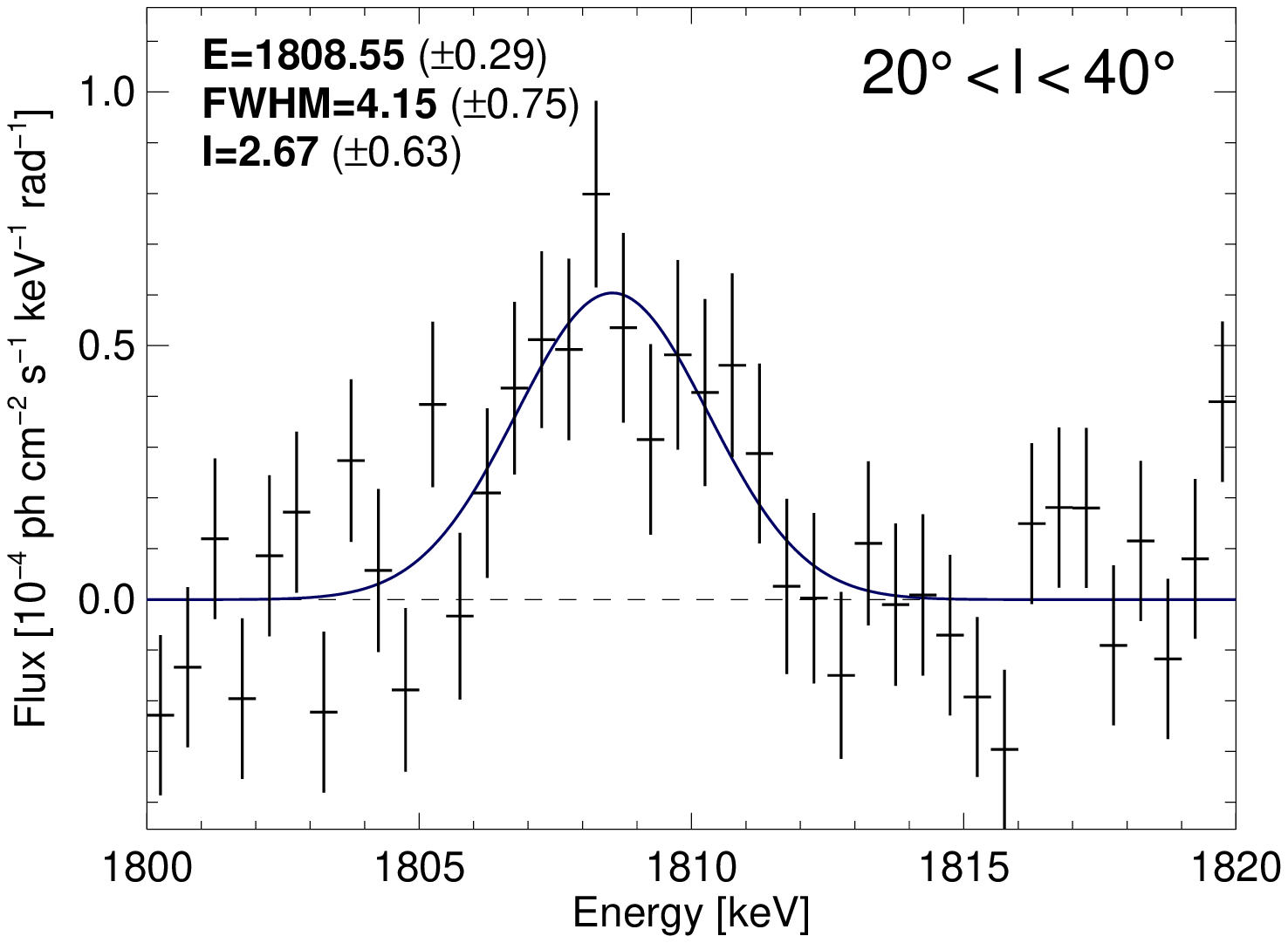}
\includegraphics[angle=0,width=5.5cm]{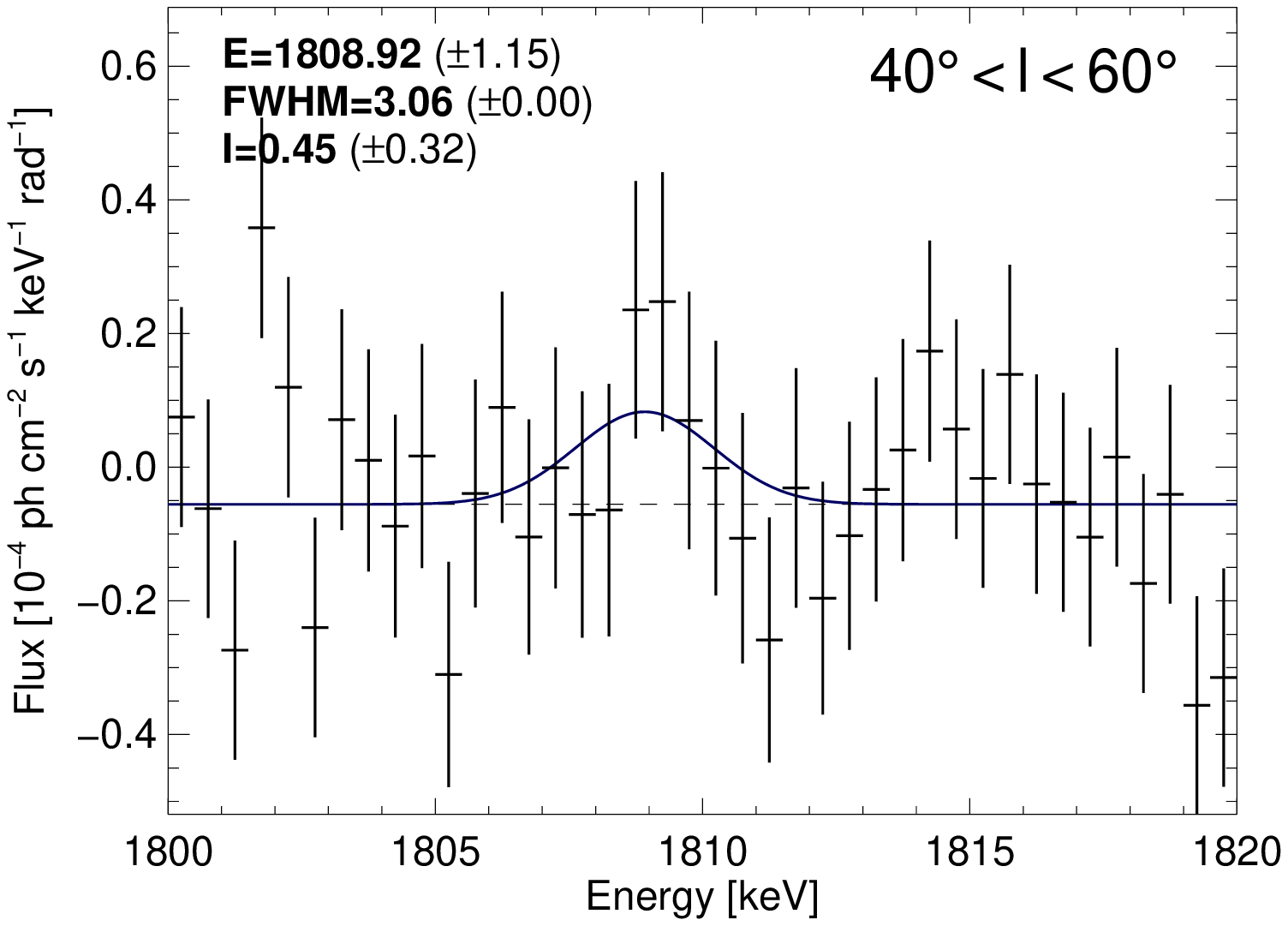}
\includegraphics[angle=0,width=5.5cm]{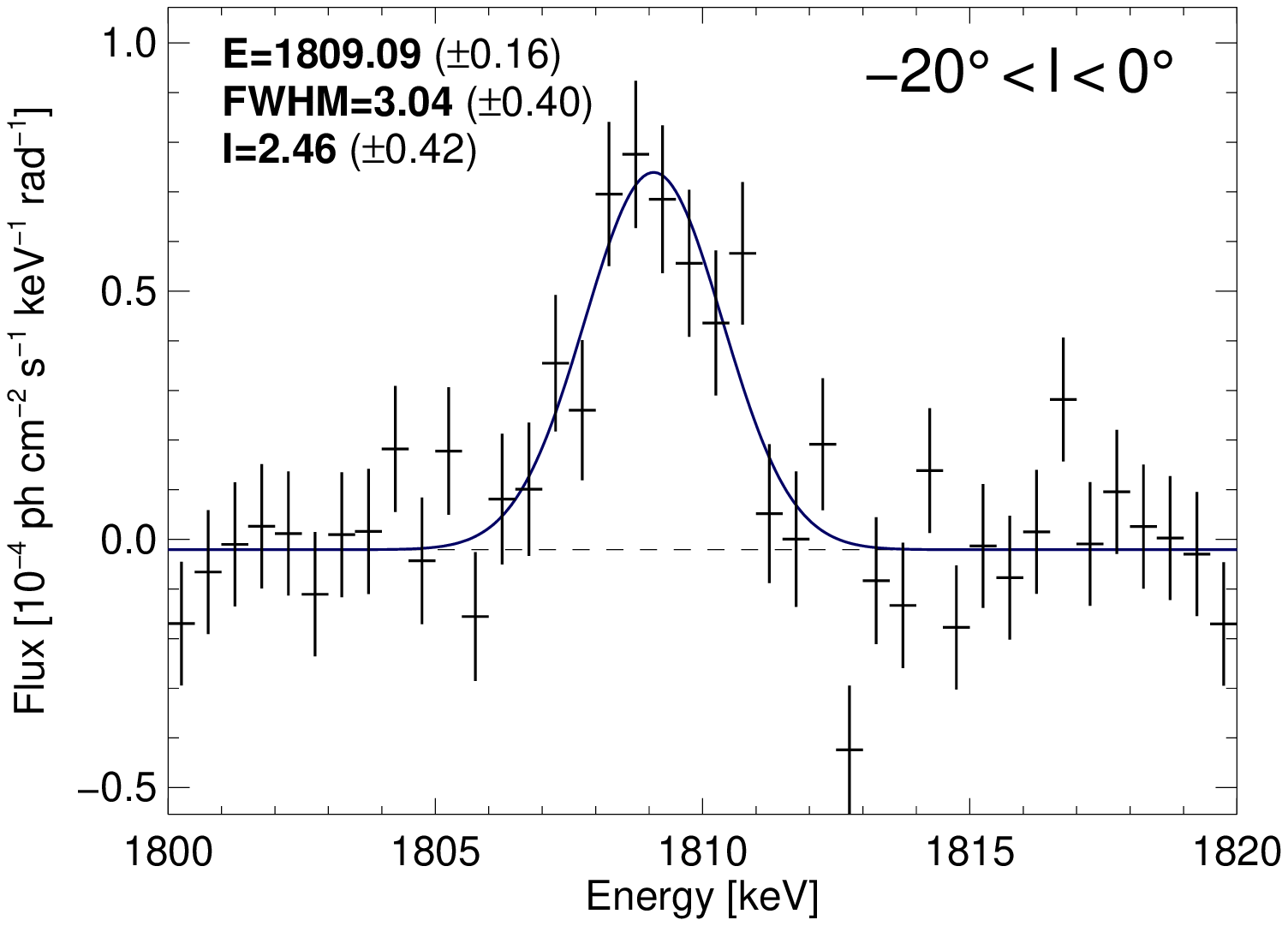}
\includegraphics[angle=0,width=5.5cm]{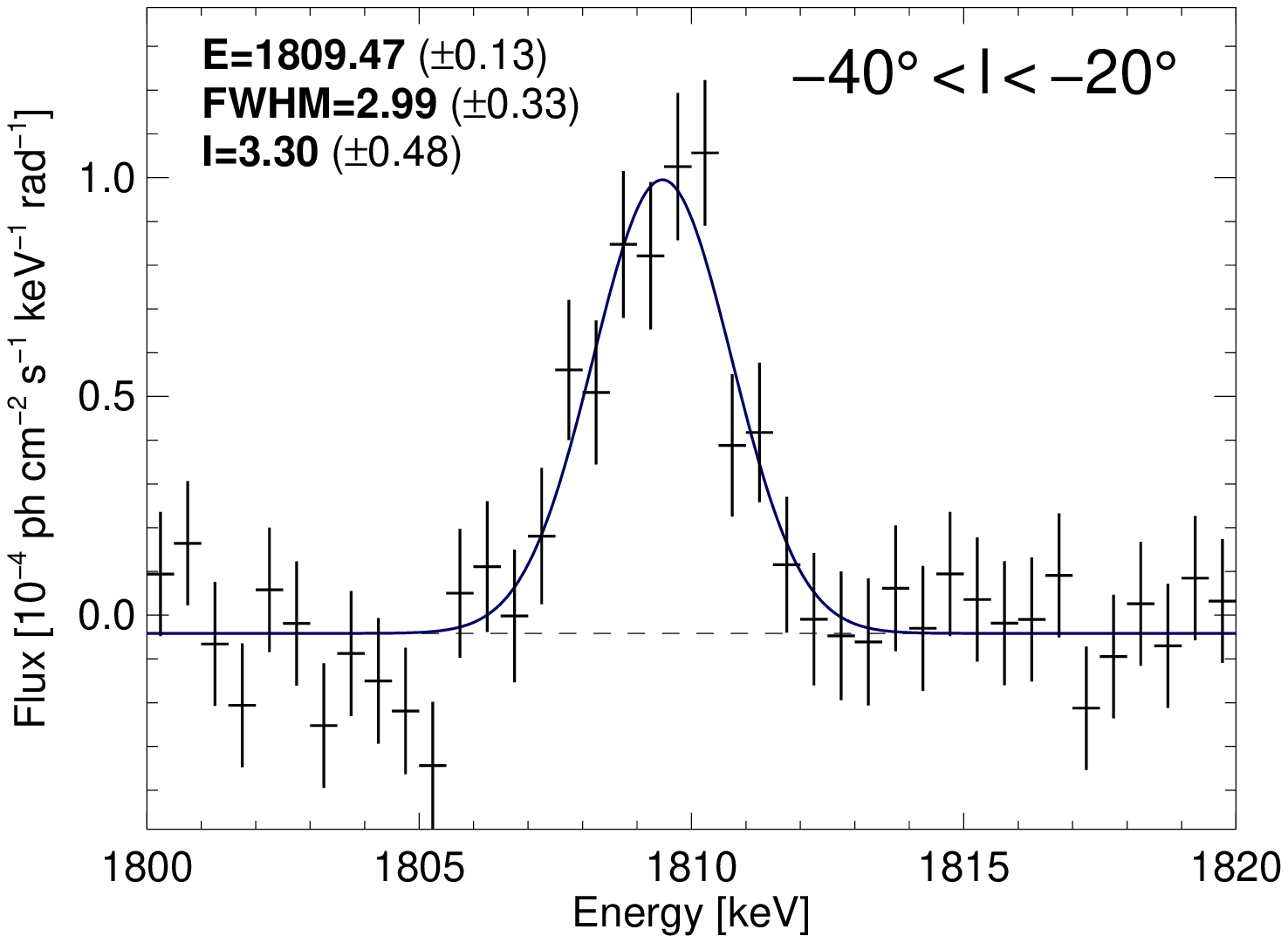}
\includegraphics[angle=0,width=5.5cm]{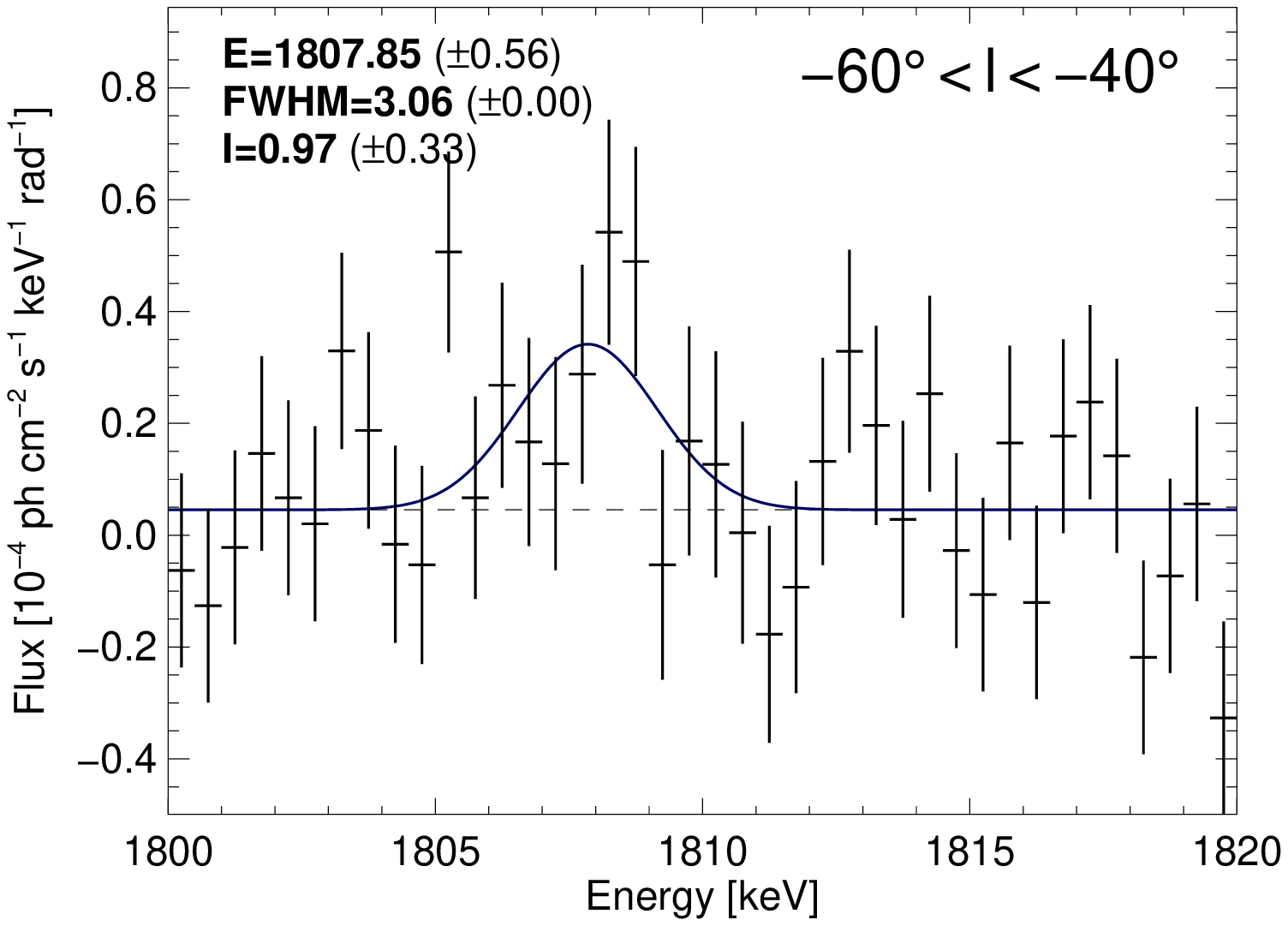}
\caption{\Al spectra of six segments along the Galactic plane (
$-60^\circ < l < 60^\circ$). Small longitude degree bin
($20^\circ$) makes the detections of \Al not significant in the
regions of $40^\circ < |l| < 60^\circ$.}
\end{figure*}

Then we also derive spectra for six smaller longitude intervals of
20 degree width along the Galactic plane ($-60^\circ < l <
60^\circ$, see Figure 10). The \Al line is still detected for the
inner Galaxy ($-40^\circ < l < 40^\circ$, $> 6\sigma$ for each
$20^\circ$ bin region), but only marginal for the two outer
regions ($40^\circ < l < 60^\circ$ and $-60^\circ < l <
-40^\circ$, $< 4\sigma$). This may be attributed to both less \Al
brightness and to less exposure in these regions, compared to the
inner Galaxy.

For the four inner regions, \Al line centroid shifts are observed.
When we use the \Al line centroid energy of 1808.65 keV from the
laboratory as a reference (rather than our own determination, Fig.
9), we obtain \Al line centroid energy shifts along Galactic
longitudes as shown in Figure 11. Evidently, towards positive
longitudes of $\sim 0^\circ - 40^\circ$, the redshift in energy is
minor, $\sim 0.1$ keV, while for negative longitudes, the
blueshift is substantial, $\sim 0.4$ keV for $-20^\circ <l<
0^\circ$ and up to 0.8 keV for $-40^\circ <l< -20^\circ$. This
asymmetry of \Al line energy shifts along the Galactic plane is
inconsistent with simple azimuthally-symmetric Galactic rotation;
it is clear evidence for pronounced spiral-arm structure in the
inner Galaxy, and possibly peculiar bulk motion along the Galaxy's
bar (e.g. Englmaier \& Gerhard 1999, Hammersley et al. 2007 and
references therein).

Figure 11 shows the intensity distribution of \Al emission (from
the same 20$^{\circ}$ regions, Figure 10) along the Galactic
plane, adding the (longitude-range normalized) Galactic-Center
region \Al intensity  ($|l|< 5^\circ$, Figure 9) for comparison.
The variability of \Al intensity along the Galactic plane again is
evident.

We note that the \Al line in the region of $20^\circ < l <
40^\circ$ appears somewhat broadened, with a Gaussian width of
FWHM$\sim 4.15\pm 0.75$ keV (also see Figure 11). This may hint
towards a peculiar \Al source region towards this direction, which
could be associated with the Aquila region (Rice et al. 2006).
Broadening could result from higher turbulence if the \Al source
region is younger than average and dominated by the \Al ejection
from more massive stars (see Kn\"odlseder et al. 2004). Further
studies would be interesting, and have the potential to identify
star formation otherwise occulted by foreground molecular clouds.

\section{Latitudinal variations of \Al emission}

The interpretation of \Al imaging and spectral results relies on
(uncertain) distances of \Al sources. Along the line-of-sight, the
detected \Al signal could originate from local star-formation
complexes ($\sim 100$ pc), or from the nearest part of the
Sagittarius-Carina arm ($1-2$ kpc), or from the Galactic center
region ($\sim 8$ kpc), or even from the distant side of the Galaxy
($> 10$ kpc).

\begin{figure}
\centering
\includegraphics[angle=0,width=7.5cm]{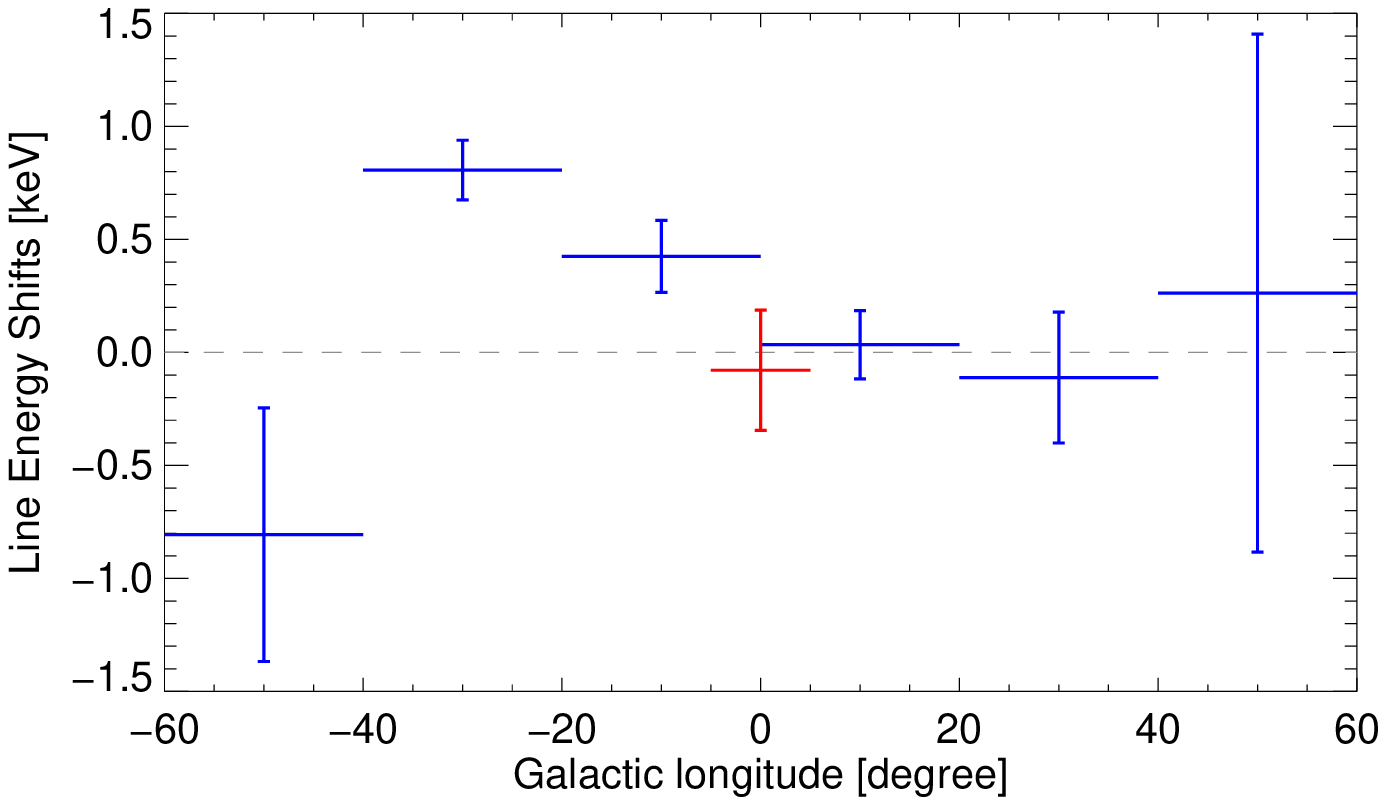}
\includegraphics[angle=0,width=7.5cm]{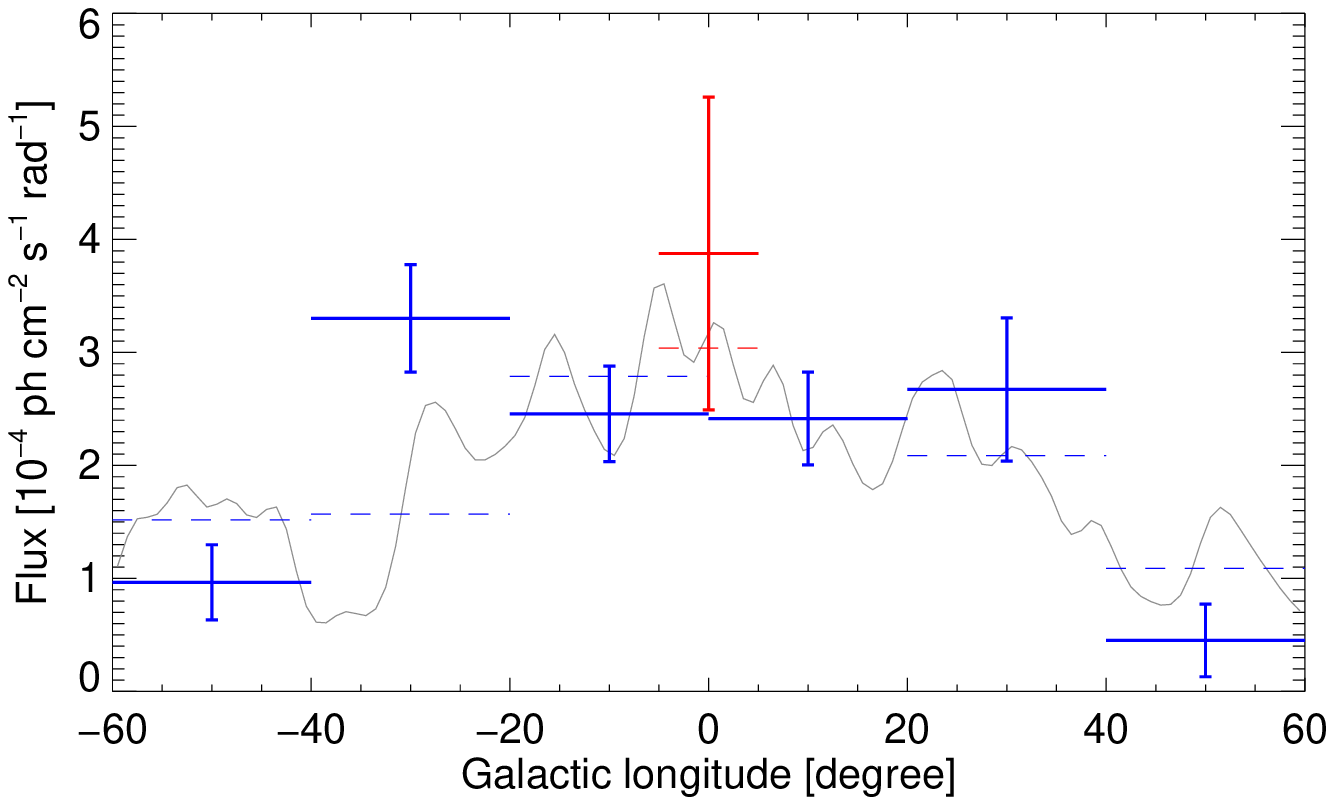}
\includegraphics[angle=0,width=7.5cm]{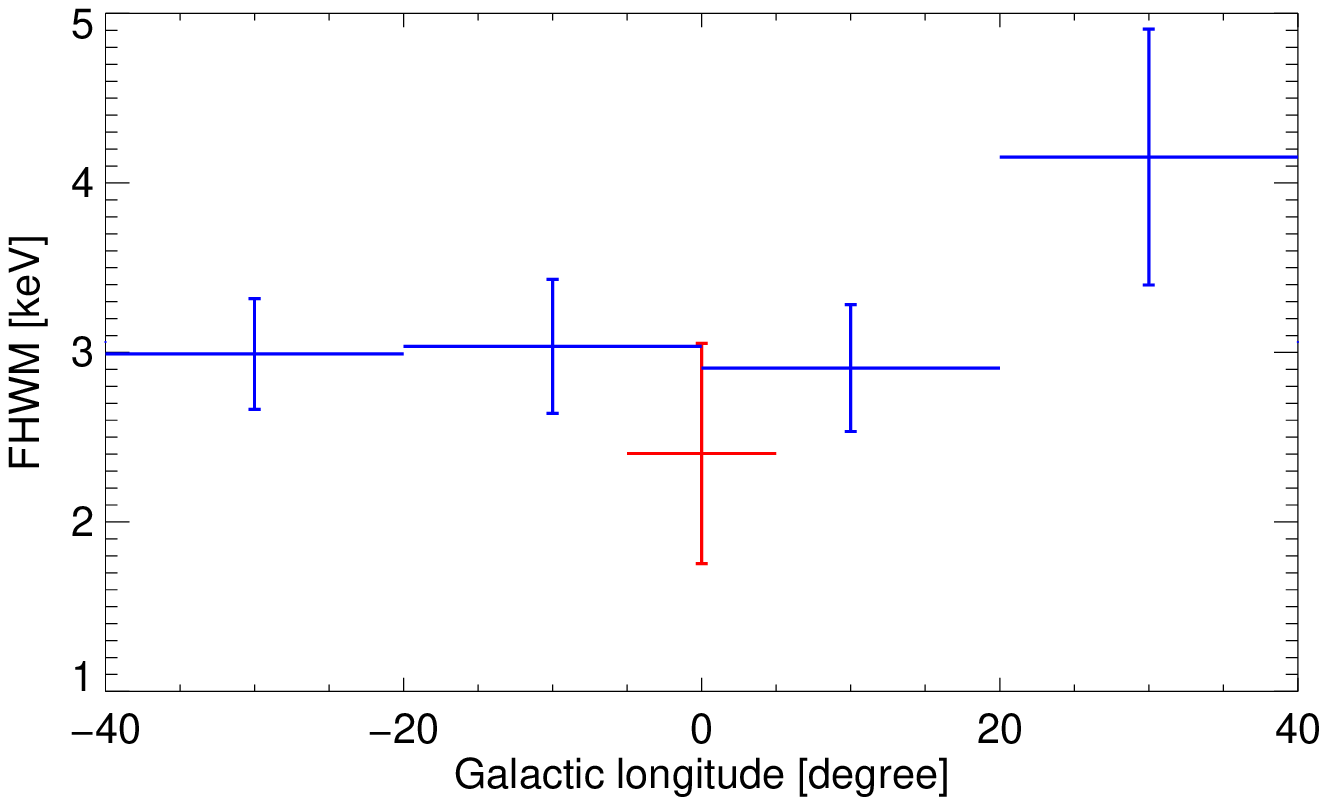}
\caption{{\bf Top:} \Al line energy shifts along the Galactic
plane ( $-60^\circ <l<60^\circ$, with longitude bin widths of
$20^\circ$, \Al spectra from Figure 10) relative to the line
centroid of the \Al line (fitted energy 1808.65 keV). {\bf
Center:} \Al intensity distribution along the Galactic plane (from
Figures 9, 10). For comparison, the COMPTEL-derived \Al intensity
profile is shown (solid line, and dashed lines when integrated
over the same longitude bins).  {\bf Bottom:} \Al FWHM (Gaussian
fitting) variation along the Galactic longitudes, a broad \Al line
feature is detected toward the longitudes $20^\circ < l <
40^\circ$. }
\end{figure}

In this section, we explore a possibility to resolve the \Al
signals for local complexes from the large scales of the Galaxy
through different latitudinal signatures. For such analysis, we
split a homogenous-disk model (Figure 3 right) into sub-maps along
Galactic latitudes, and derive separate \Al spectra for the
different latitude ranges simultaneously through model fitting to
our observations. We study the \Al emission for three intervals
along latitudes, $-5^\circ<b<5^\circ $ (low latitudes),
$5^\circ<b<20^\circ$ and $-20^\circ<b<-5^\circ$ (intermediate
latitudes). \Al emission for low latitudes ($|b|<5^\circ$) should
be dominated by a large-scale origin in the Galactic disk, while
\Al emission at higher latitudes ($|b|>5^\circ$) could originate
from more local star-formation systems such as OB associations in
the Gould Belt. This definition is similar to the one used in
pulsar population studies (Wang et al. 2005). The Gould Belt is an
ellipsoidal shaped ring-like structure delineated by the groups of
nearby stellar groups within \about~1~kpc, with semi-major and
minor axes of $\sim 500$ pc and 340 pc, respectively (Perrot \&
Grenier 2003). The center of this structure is located towards
$l=130^\circ$ , displaced from the Sun's location by about 200 pc
(Guillout et al. 1998). The Vela region is located near the outer
boundary of the Gould Belt towards $l\sim -90^\circ$ . The nearby
Sco-Cen region at about 140~pc distance towards $l\sim -10^\circ,
\ b\sim 10^\circ$ probably also belongs to the Gould Belt. The
origin of this structure is debated, between triggered sequential
star formation propagating outwards from its center, and an
external triggering event such as a high-velocity cloud falling
through the Galaxy's plane (see Perrot \& Grenier 2003 and
P\"oppel 1997).

Figure 12 displays the \Al intensity distribution along Galactic
latitudes for $|l|<60^\circ$. No \Al signal is detected at
negative latitudes ($-20^\circ<b<-5^\circ$), while weak \Al
emission is detected at positive latitudes ($\sim 2\sigma,\
5^\circ<b<20^\circ$). We also derive latitudinally-separated \Al
emission for the 1st and 4th quadrants separately again through
model fitting, also shown in Figure 11. No \Al signal is detected
for off-plane regions of the 1st quadrant, while in the 4th
quadrant, \Al emission is still clearly detected (3$\sigma$) in
the latitude region $5^\circ<b<20^\circ$. This \Al emission
towards $b>5^\circ,\ l<0^\circ$ could be attributed to the nearby
Sco-Cen star-formation complex at \about~140~pc distance.

We also determine the scale height of the Galactic plane in \Al
emission by comparing the fit quality for sets of two different
plausible geometrical models for the \Al source density
distribution in the Galaxy, varying their scale height parameter
(Fig. 12 right). We use a Galactocentric double-exponential disk
model (ExpDisk) described by
\begin{equation}
\rho(R,z)\propto\mathrm{e}^{-\left( \frac{R}{R_0} +
\frac{|z|}{z_0} \right)},
\end{equation}
where $R$ is the Galactocentric distance within the plane, and $z$
is the height above the plane, with $R_0=3.5$~kpc and $z_0$ as the
scale radius and height parameter. Alternatively, we use a model
which includes the spiral-arm structure of the Galaxy as derived
by Taylor \& Cordes (1993); we use only their components for the
inner Galaxy and the spiral arms, with identical scale height
parameter (in this model, the density perpendicular to the disk is
described as sech$\left(z/z_0\right)$). In order to avoid a bias
from bright special regions such as Cygnus/Vela/Carina, we
restrict this analysis to data within $|l|<60\degr$ and
$|b|<30\degr$. Fig. 12 (right) shows the variation of
log-likelihood values with different scale heights for both model
types. Here, the values for the exponential-disk models have been
shifted by $+16$, as the spiral-arm model systematically provides
a better description of our data. With $-2\log$L being
asymptotically $\chi^2$ distributed, we derive a scale height of
$130^{+120}_{-70}$~pc $(1\sigma)$ for the \Al emission in the
inner Galactic disk, from the spiral arm model constraints. This
confirms previous such studies based on COMPTEL data (Diehl et al.
1998).

\begin{figure*}
\centering
\includegraphics[width=8.5cm]{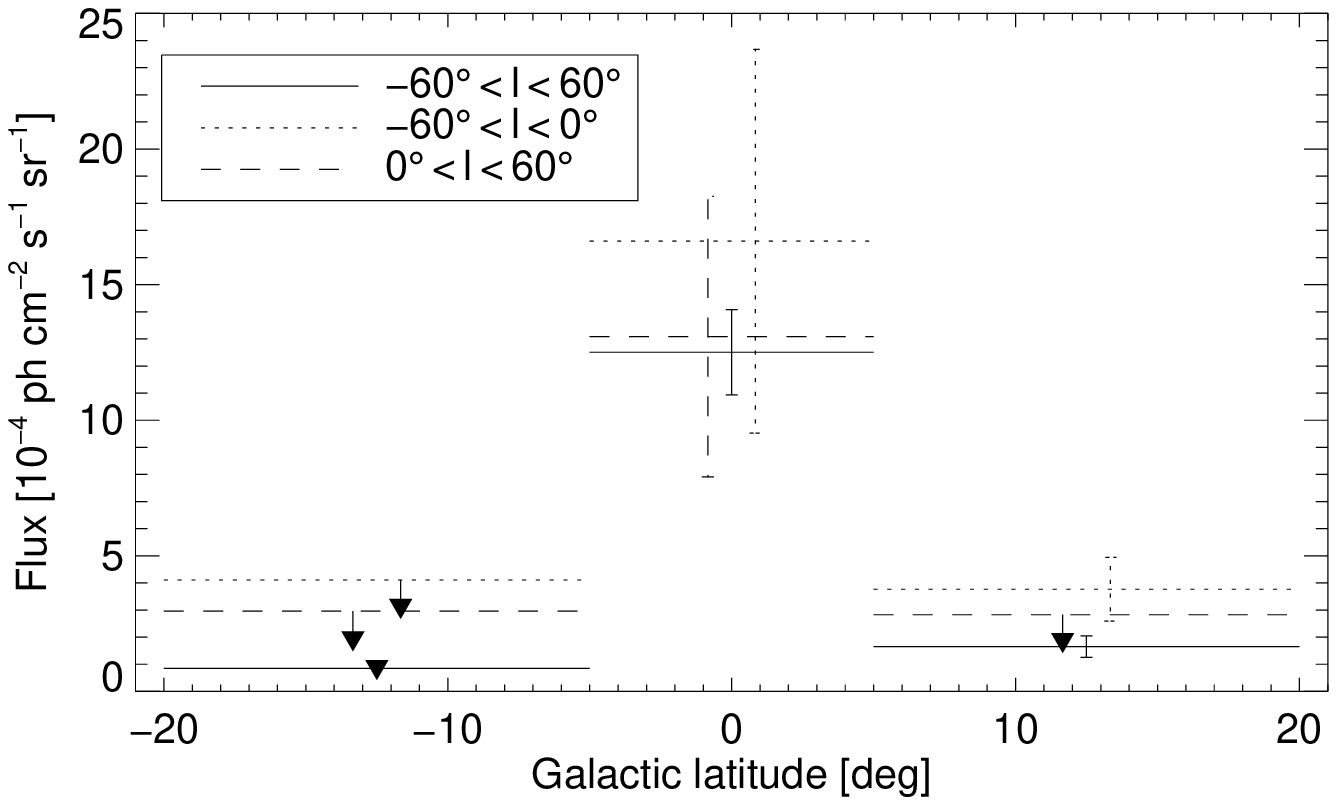}
\includegraphics[width=8.5cm]{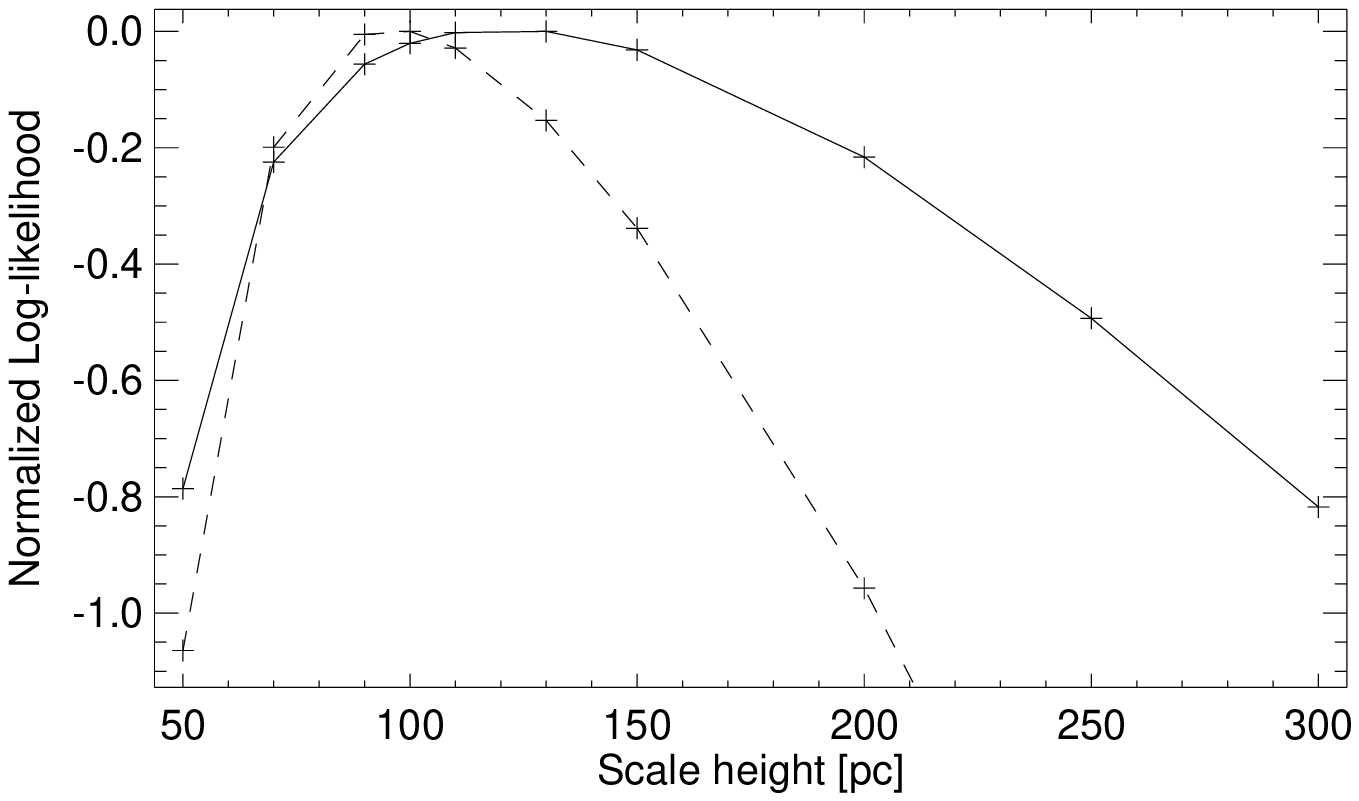}
\caption{\textbf{Left:} \Al intensities versus Galactic latitudes
(solid lines, $|l|<60^\circ$). \Al emission from low latitudes
(the Galactic disk) dominates ($|b|<5^\circ$). For comparison, \Al
intensities for each Galactic quadrant are also shown. Significant
\Al emission is detected towards $b>5^\circ,\ l<0^\circ$.
\textbf{Right:} Determination of the Galactic-disk scale height.
Shown is the logarithmic likelihood-ratio for the exponential-disk
(dashed line) and spiral-arm structure models (solid line) for
different scale height values. Log-likelihood values for the
exponential-disk model have been offset by $+16$. }
\end{figure*}

\section{Summary and Conclusions}

The study of the \Al line and its details in different regions of
the Galaxy is one of the main goals of the SPI spectrometer on
INTEGRAL. Using five years of SPI data, we find the \Al signal
from the inner Galaxy ($|l|<30^\circ,\ |b|<10^\circ $) with a high
significance of $\sim 28\sigma$. The \Al flux integrated over the
inner Galaxy of $(2.9\pm 0.2)\times 10^{-4}\ \mathrm{ph\ cm^{-2}\
s^{-1}\ rad^{-1}}$ is consistent with our own and other earlier
measurements, though significantly lower than the ones from
measurements with instruments which have even more modest spatial
resolution on the sky (see Figure 2 of Diehl et al. 2004). Taking
the distance of the Sun to the Galactic center as $R_0=$ 8.5 kpc,
we convert the measured \Al flux to a Galactic \Al mass of
$(2.7\pm 0.7)$ \ms. We compared different plausible models for the
sky distribution of \Al emission and how it may affect the global
\Al flux measurement. We find that different sky distribution
models do not substantially affect \Al intensity and line shapes
for the integrated inner Galaxy, and we have used the observed
variations of \Al mass values to estimate a systematic
uncertainty, added to the statistical uncertainty from the model
fitting, to yield the $\pm 0.7$ \ms uncertainty quoted.

The \Al line centroid energy appears blue-shifted relative to the
laboratory value of $1808.65\pm 0.07$ keV if integrated over the
inner Galaxy ($1809.0\pm 0.1$ keV). With refined spatial
resolution this turns out to be mostly due to asymmetric bulk
motion which we find along the plane of the Galaxy, and attribute
to asymmetries in inner spiral arms and to the Galaxy's bar. In
particular, for the central longitude bin towards the Galactic
Center, the signal is strong enough to allow for a rather small
longitude range integration, and we find a \Al line centroid
consistent with the laboratory value and thus with the absence of
bulk motion relative to the Sun. Also, here the \Al line appears
as most-narrow with an upper limit of 1.3 keV (2$\sigma$). The
measured line width of \Al from the large-scale integrated inner
Galaxy with an upper limit of 1.3 keV (2$\sigma$) is consistent
with expectations from both Galactic rotation and modest
interstellar-medium turbulence around the sources of \Al
(turbulent velocities constrained below 160 km s$^{-1}$ even when
disregarding the effects of galactic rotation).  Our line width
results are consistent with previous reports by HEAO-C (Mahoney et
al. 1984), RHESSI (Smith 2003), but the very broad line with a
width $\sim 5.4$ keV reported by GRIS (Naya et al. 1996) is
clearly ruled out by our SPI measurements ( $4\sigma$ significance
level ).

From our study of \Al emission in spatially-restricted regions along the
plane of the Galaxy, we find:
\begin{itemize}
\item{}
\Al brightness appears asymmetric for the two inner quadrants,
with a flux ratio of the 4th quadrant to the 1st of $\sim 1.3 \pm
0.2$ (Figure 7).
\item{}
The \Al line energy varies clearly along the Galactic plane
(Figure 11): a minor redshift ($\sim 0.1$ keV) for positive
longitudes, but significant blueshift ($\sim 0.4-0.8$ keV) for
negative longitudes.
\item{}
The \Al line towards the direction of $20^\circ < l < 40^\circ$
shows a hint for additional line broadening.
\item{}
The scale height of Galactic-plane \Al emission is
$130^{+120}_{-70}$~pc $(1\sigma)$, determined towards
the inner Galaxy ($|l|<60\degr$).
\item{}
There is a strong hint for \Al emission in the fourth Galactic
quadrant at intermediate latitudes $l<0^\circ,\ b>5^\circ$ (Fig.11).
\end{itemize}

This leads us to several astrophysical implications and the
``bar'' structures:
\begin{itemize}
\item{}
The disk of our Galaxy apparently is not azimuthally-symmetric.
The \Al brightnesses for the two inner quadrants appear different,
both from our INTEGRAL and from earlier COMPTEL results, favoring
the fourth quadrant. \Al line shifts towards the blue in such
brighter regions are more pronounced than the redshifts on the
fainter side of the disk, unexpected from a simple and symmetric
Galaxy model and its large-scale rotation properties. These
phenomena may reflect variations in the space distribution of
young stars, possibly related to the inner parts of the spiral
arms and their interfaces to the ``molecular ring'' and the
``bar'' structures. Large non-circular motions have been seen in
HI and CO observations (Mulder \& Liem 1986; Gerhard \& Vietri
1986), and also from the NIR light distribution (Binney et al.
1997), in addition to source count asymmetries (Nikolaev \&
Weinberg 1997) and non-symmetric gas dynamics (Englmaier \&
Gerhard 1999). The bar is most clearly traced in infrared emission
from dust (Marshall et al. 2008), which suggests a position angle
of $\sim 25^\circ$ of the bar with the near end pointing towards
us in the first quadrant, and a total length of about 4~kpc. The
transition regions between spiral arms and bar are likely to incur
star formation (Verley et al. 2007), and affect the dynamics of
gas and stars. Simply superimposing a homogeneous star-forming bar
to the rotating spiral would, however, not agree with our results
- the inner Galaxy structure may be more complex.  Localized
star-forming regions could lead to peculiar motion of hot gas
ejected from winds and supernovae, which might dominate over
large-scale Galactic rotation. Such activity could be the cause of
the observed asymmetry of the \Al line intensities and line energy
shifts in the inner Galaxy. On the other hand, other more nearby
\Al source regions could be responsible for these irregularities,
such as the nearest part of the Sagittarius-Carina arm in the
fourth quadrant, or regions/complexes attributed to the Gould Belt
such as the Scorpius-Centaurus-Lupus groups. Refined \Al studies
and their combination with astrometry from other tracers of
Galactic structure could help to understand the role of our
Galaxy's bar.
\item{}
The \Al emission asymmetry ($1.3\pm 0.2$) between fourth and first
quadrant of the Galaxy is lower than the intensity contrast of
$1.8^{+0.5}_{-0.3}$ reported for positron annihilation emission
from the disk of the Galaxy (Weidenspointner et al. 2008). \Al by
itself releases a positron in 82\% of its decays; it is uncertain,
however, how this translates into annihilation photons, from the
variety of slowing down and annihilation processes which determine
the fate of positrons in interstellar space. Both spatial and
temporal variations and non-linearities scaling with gas density
may occur.
\item{}
The scale height of Galactic-plane \Al emission is significantly
larger than the molecular-gas
disk scale height of \about~50~pc, yet significantly smaller than
the ``thick disk'' part of the Galaxy. It is consistent with
\Al being ejected from star-forming regions, and partly extending
more towards the Galactic halo where gas pressure is lower than
within the plane of the Galaxy (``champagne flows'').
\item{}
Localized regions may deviate in interesting detail from the
large-scale averaged properties of \Al source regions. The
direction of $20^\circ < l < 40^\circ$ corresponds to the Aquila
region, and our hint for additional line broadening may be due to
increased interstellar turbulence from stellar-wind and supernova
activity at the characteristic age of stellar groups of that
region, which may have created a supershell of substantial size
(320 $\times$ 550~pc, see Maciejewski et al. 1996). Similar
arguments are being explored for the Cygnus region, based on
earlier hints of peculiar \Al emission (Martin et al. in
preparation). The hint for \Al emission at $l<0^\circ,\ b>5^\circ$
(Fig. 12) may be attributed to the relatively nearby star forming
complexes of the Sco-Cen association (de Geus 1992). Sco-Cen and
several nearby stellar groups are attributed to the larger
structure of the ``Gould Belt'', which is suggested to have been
more actively forming stars during the last 30 Myr (Grenier 2000,
Perrot \& Grenier 2003). These examples indicate that
spatially-resolved \Al emission properties may enable new
diagnostics of the interactions of massive stars with their
surroundings, when combined with other astronomical constraints on
cold gas and stars of such regions.
\end{itemize}

INTEGRAL/SPI will continue to accumulate more data for several
years, to cover more sky regions in the outer Galaxy and at
intermediate latitudes. This will allow us to deepen the studies
reported in this paper, and to refine spatial information on the
\Al line and/or increase the significance of the results reported
above, as signal statistics increases and instrumental-background models
are tightened.   These studies will help us to better understand the
properties of ISM near groups of massive stars, and the bulk
motion of gas in the inner Galaxy from Galactic rotation and other
peculiar kinematics. Nearby \Al sources may be also discriminated
with better exposure towards candidate regions, improving our
determination of the \Al mass in the Galaxy and towards regions where the
stellar census is known to a better degree.
Studies of \Al from nearby star-formation
regions (e.g., Cygnus, Vela, Sco-Cen, Orion) are a promising diagnostic
for the massive star origin of Galactic \Al and kinematics of \Al
ejecta in ISM. From this, important constraints on massive-star
and supernova nucleosynthesis are obtained, refining our models of
their complex interiors (see e.g. Woosley and Heger 2007, or
Chieffi and Limongi 2006).

\section*{Acknowledgments}
We are grateful to the anonymous referee for the comments to
improve the draft. The INTEGRAL project is supported by government
grants in the member states of the hardware teams. The SPI project
has been completed under responsibility and leadership of CNES. We
are grateful to ASI, CEA, CNES, DLR, ESA, INTA, NASA, and OSTC for
support. W. Wang is also supported by the National Natural Science
Foundation of China under grant 10803009.

{}

\end{document}